\documentclass[english]{article}
\usepackage[T1]{fontenc}
\usepackage{amsmath}
\usepackage{amssymb}
\usepackage{graphicx}
\usepackage{caption}
\newtheorem{thm}{Theorem}%[section]
\newtheorem{proof}{Proof}
\makeatletter
\@ifundefined{showcaptionsetup}{}{%
 \PassOptionsToPackage{caption=false}{subfig}}
\usepackage{subfig}
\makeatother
\usepackage{babel}
\begin{document}

\title{Bright-dark soliton solutions to the multi-component AB system}
\author{Zong-Wei Xu, Guo-Fu Yu\footnote{Corresponding author. Email address: gfyu@sjtu.edu.cn}
and Zuo-Nong Zhu\\
School of Mathematical Sciences, Shanghai Jiao Tong University, \\
Shanghai 200240, P.R.\ China}
\date{}
\maketitle
\begin{abstract}
In this paper we investigate the multi-component AB system that comes from the geophysical fluid dynamics. We construct its bright-dark soliton solutions through Hirota's bilinear method. For the two-component AB system, asymptotic behaviours of two-soliton solution are obtained and interactions between two bright and two dark solitons are proved to be elastic. Under different parameter conditions, the oblique interactions, bound states of solitons are analyzed in details. Meanwhile, by use of the Pfaffian technique, we present $N$-bright and $N-$dark soliton solutions to the two- and multi-component AB system. The results will be meaningful for the study of vector multi-dark solitons in many physical systems such as nonlinear optics and fluid dynamics.
\end{abstract}

{\bf keywords}: AB system; bright-dark soliton; Pfaffian; Hirota method \vskip 5pt

%{\bf
%The AB system describes  two-layer marginally stable
%geophysical flows near the neighborhood of a appropriate vertical
%shear. The system attracts great attention and  many integrable properties together with its rich solutions  have been reported. In this paper, we consider the coupled AB system. We construct bright-dark soliton solutions to the two-component AB system. Asymptotic behaviours of two-soliton solution are obtained and interactions between two bright and two dark solitons are proved to be elastic. Under different parameter conditions, the oblique interactions, bound states of solitons are analyzed in details. The compact $N-$bright-dark-soliton solutions to the coupled AB systems are presented by use of the Pfaffian.
%}
\section{Introduction}

The AB system
\begin{align}
 & \left(\partial_{T}+c_{1}\partial_{X}\right)\left(\partial_{T}+c_{2}\partial_{X}\right)A=n_{1}A-n_{2}AB_{0},\label{ab1}\\
 & \left(\partial_{T}+c_{2}\partial_{X}\right)B_{0}=\left(\partial_{T}+c_{1}\partial_{X}\right)\left|A\right|^{2},\label{ab2}
\end{align}
is a completely integrable model for two-layer marginally stable
geophysical flows near the neighborhood of a appropriate vertical
shear \cite{ab1,ab2}. Here $A(X,T)$ is a complex valued wave packet and $B_{0}(X,T)$ describes
the motion of the basic flow induced by the wave packet. Parameters $c_1,c_2$ and $n_1,n_2$ are arbitrary real constants. Although the AB system was first found to be a geophysical fluid model, it can describe the ultrashort optical pulse propagation in nonlinear optics \cite{NW} and the mesoscale gravity current transmission in the problem of cold gravity current \cite{gcurr}. Solutions, including breathers \cite{periodic-solu,ab3,Breathers}, rogue waves \cite{GDT1,GDT2} and  modulation instability \cite{chow}, bright and dark solitons \cite{Yu,Yan}, and integrable aspects \cite{tianbo} to the AB system have drawn widespread attention in recent decades.

Through the dependent variable transformation
\begin{equation}
x=X-c_{1}T,\quad t=T-\frac{X}{c_{2}},\quad B_{0}=\frac{(c_{1}-c_{2})^{2}}{n_{2}c_{2}}B,\label{eq:DVP_C}
\end{equation}
the AB system \eqref{ab1}-\eqref{ab2} transforms into the canonical form
\begin{align}
A_{xt}=\mu A-AB,\quad B_{x}=\sigma (|A|^2)_t,\label{ab}
\end{align}
with $\mu=\frac{n_1c_2}{(c_1-c_2)^2}$ and $\sigma=\frac{n_2}{(c_1-c_2)^2}$. The compatibility condition of the system \eqref{ab} yields
\begin{align}
2\sigma |A_t|^2+B^2-2\mu B=f(t),
\end{align}
where $f(t)$ is a function of integration. The parameter $\mu>0$ or $\mu<0$ denotes supercritical and subcritical cases of the shear, respectively, and $\sigma$ denotes the nonlinearity. When $\sigma$ is positive, the AB system could be reduced to sine-Gordon equation. Reversely, if $\sigma$ is negative, the AB system could be transformed into single sinh-Gordon equation.
Upon a shift $\mu-B\rightarrow B$ and new symbol $\gamma=-\sigma$, the AB system transforms into a simpler form
\begin{align}
A_{xt}=AB,\quad B_x=\gamma (|A|^2)_t. \label{s-ab}
\end{align}
In ref. \cite{Yan}, the Darboux transformation for \eqref{s-ab} with $\gamma=\frac{1}{2}$ via
the loop group method is constructed and associated $N$-fold Darboux transformation is found in terms of simple
determinants. As a bonus, multi-dark-dark solitons of the AB system \eqref{s-ab} with a
non-vanishing background  are obtained.

Different from Darboux transformation method, dark soliton solutions can also be arrived by Hirota's bilinear method. When $\sigma>0$, through the dependent variable transformation
\begin{align}
A=\frac{g}{f},\quad B=2(\ln f)_{xt},\quad g\,\, \mbox{complex},\, f\,\,\mbox{real}, \label{vb1}
\end{align}
we can rewrite Eq.\eqref{ab} into bilinear form
\begin{align}
&D_{x}D_{t}f\cdot g=\mu fg,\label{b-ab-1}\\
&D_{x}^{2}f\cdot f=\sigma |g|^{2}, \label{b-ab-2}
\end{align}
where $D$ is the Hirota operator defined as
\begin{align}
D_x^nD_t^m f\cdot g=\frac{d^n}{d\epsilon^n}\frac{d^m}{d\delta^n}\Big(f(x+\epsilon,t+\delta)g(x-\epsilon,t-\delta)\Big)\Big|_{\epsilon=0,\delta=0}.
\end{align}
Upon  the reduction of two-component extended KP hierarchy, the $N-$bright soliton solution in the Gram determinant form for the AB system \eqref{ab} was found \cite{Yu}.

When $\sigma<0$, we take the following dependent variable transformation
\begin{align}
A=\sqrt{\frac{-1}{2\sigma}}\frac{g}{f}e^{\mathrm{i}(t-x)},\quad B= \mu-1+2 (\ln f)_{xt},
\end{align}
and rewrite the AB system \eqref{ab} into bilinear equations
\begin{align}
&(D_xD_t +\mathrm{i} D_t-\mathrm{i} D_x )f\cdot g=0,\label{d-ab-1}\\
& (D_x^2-\frac{1}{2}) f\cdot f=- \frac{1}{2}gg^*,\label{d-ab-2}
\end{align}
where $^*$ means complex conjugate. Based on the Sato theory and reduction technique, the $N-$dark soliton solution in the Gram determinant form was constructed \cite{Yu}.

It is well known that in the study of the propagation of optical pulses in birefringence fibers or anisotropy effect, the  multi-component systems should be considered. Compared with the single-component case, the multi-component ones own richer properties. Owing to the different polarization of each component, multi-component system can display bright-bright, bright-dark and dark-dark solutions.

The $M$-component AB system in the canonical form
\begin{align}
 & A_{k,xt}=\mu A_{k}-A_{k}B,\;k=1,2,\cdots,M\label{eq:mc1},\\
 & B_{x}=\sigma\left(\sum_{k=1}^{M}\left|A_{k}\right|^{2}\right)_{t},\label{eq:mc2}
\end{align}
associated with Lax pair has been presented in \cite{Yu}.
In the $X-T$ coordinates, the $M$-component  AB equation
is in the form
\begin{align}
 & \left(\partial_{T}+c_{1}\partial_{X}\right)\left(\partial_{T}+c_{2}\partial_{X}\right)A_{k}
 =n_{1}A_{k}-n_{2}A_{k}B_{0},\;k=1,2,\cdots,M\label{eq:eqm1}\\
 & \left(\partial_{T}+c_{2}\partial_{X}\right)B_{0}
 =\left(\partial_{T}+c_{1}\partial_{X}\right)\sum_{k=1}^{M}\left|A_{k}\right|^{2}.\label{eq:eqm2}
\end{align}

To the best of our knowledge, the bright-dark vector soliton solutions and their interactions for coupled AB systems have not been put forward. Motivated by the above reason, in this paper, we shall study the bright-dark vector soliton solutions and their interactions for eqs. \eqref{eq:mc1}-\eqref{eq:mc2} via the Hirota method. The rest of the paper is organized as follows. In Section 2, the bright-dark one-,two- and $N-$soliton solutions to the two-component AB system will be constructed
based on the bilinear equations. In Section 3,  asymptotic analysis and oblique interaction will be performed on the two-soliton solutions. Bound states of solitons and periods will be analyzed in Section 4. $N$-bright-dark soliton solutions to the multi-component AB system in the Pfaffian form are derived in Section 5.  Conclusion is given in Section 6.

\section{Bright-dark soliton solutions to the two-component AB system}
Setting $M=2$ in \eqref{eq:eqm1}-\eqref{eq:eqm2}, we have the two-component AB system
\begin{align}
 & \left(\partial_{T}+c_{1}\partial_{X}\right)\left(\partial_{T}+c_{2}\partial_{X}\right)A_{1}=n_{1}A_{1}-n_{2}A_{1}B_{0},\label{eq:eq21}\\
 & \left(\partial_{T}+c_{1}\partial_{X}\right)\left(\partial_{T}+c_{2}\partial_{X}\right)A_{2}=n_{1}A_{2}-n_{2}A_{2}B_{0},\label{eq:eq22}\\
 & \left(\partial_{T}+c_{2}\partial_{X}\right)B_{0}=\left(\partial_{T}+c_{1}\partial_{X}\right)\left(\left|A_{1}\right|^{2}+\left|A_{2}\right|^{2}\right),\label{eq:eq23}
\end{align}
 and its canonical form
\begin{align}
 & A_{1,xt}=\mu A_{1}-A_{1}B,\label{eq:2c1}\\
 & A_{2,xt}=\mu A_{2}-A_{2}B,\label{eq:2c2}\\
 & B_{x}=\sigma\left(\left|A_{1}\right|^{2}+\left|A_{2}\right|^{2}\right)_{t},\label{eq:2c3}
\end{align}
where two sets of coordinates  $x,t$ and $X,T$ are related by (\ref{eq:DVP_C}).

The Hirota bilinear method is an effective way to derive multi-soliton
solutions to nonlinear evolution equations. We first implement the dependent
variable transformation
\begin{align}
A_{1}=\frac{g_{1}}{f},\thinspace A_{2}=\frac{g_{2}}{f},\thinspace B=2\left(\ln f\right)_{xt}-\lambda ,\label{vt}
\end{align}
and rewrite two-component AB system (\ref{eq:2c1})-(\ref{eq:2c3})
into the bilinear form
\begin{align}
 & D_{x}D_{t}g_{1}\cdot f=\left(\mu+\lambda\right)g_{k}f,\label{eq:b2c1}\\
 & D_{x}D_{t}g_{2}\cdot f=\left(\mu+\lambda\right)g_{2}f,\label{eq:b2c2}\\
 & \Big(D_{x}^{2}+r^{2}\sigma\Big)f\cdot f=\sigma\left(\left|g_{1}\right|^{2}+\left|g_{2}\right|^{2}\right),\label{eq:b2c3}
\end{align}
where $r$ and $\lambda$ are real parameters.

\subsection{One-soliton solution}

In this subsection, we present the one-bright-dark-soliton solution to
(\ref{eq:eq21})-(\ref{eq:eq23}). To obtain such solution, we set
$g_{1},\thinspace g_{2}$ and $f$ in the form
\begin{align}
 & g_{1}=e^{\eta_{1}},\label{eq:2c1g1}\\
 & g_{2}=re^{i\varphi}\left(1+a_{1,1^{*}}b_{1,1^{*}}e^{\eta_{1}+\eta_{1}^{*}}\right)\label{eq:2c1g2},\\
 & f=1+a_{1,1^{*}}e^{\eta_{1}+\eta_{1}^{*}},\label{eq:2c1f}
\end{align}
where
\begin{align}
 & \eta_{1}=k_{1}x+\omega_{1}t+\eta_{1,0}\, \in\mathbb{C},\label{eq:2ceta1}\\
 & \varphi=p\,x-q\,t+\varphi_{0}\,\in\mathbb{R},\label{eq:2cphi}
\end{align}
Substitution expressions (\ref{eq:2c1g1})-(\ref{eq:2c1f}) into (\ref{eq:b2c1})-(\ref{eq:b2c3}) leads to
\begin{alignat}{2}
 & a_{1,1^{*}}=\frac{1}{\left(k_{1}+k_{1}^{*}\right)^{2}\left(\frac{2}{\sigma}+\frac{4p^{2}r^{2}}{\left|p^{2}+k_{1}^{2}\right|^{2}}\right)}, & \thinspace\quad & \omega_{1}=\frac{\mu+\lambda}{k_{1}},\\
 & b_{1,1^{*}}=\frac{\left(k_{1}-ip\right)\left(k_{1}^{*}-ip\right)}{\left(k_{1}+ip\right)\left(k_{1}^{*}+ip\right)}, & \thinspace & q=\frac{\mu+\lambda}{p}.
\end{alignat}
Setting $\theta_{1}=\ln\frac{k_{1}-ip}{k_{1}+ip}=\ln\left|\frac{k_{1}-ip}{k_{1}+ip}\right|+i\arg\frac{k_{1}-ip}{k_{1}+ip}$,
we can express $b_{1,1^*}$ in compact form $ b_{1,1^{*}}=\exp(\theta_{1}-\theta_{1}^{*})$.
Thus from the dependent variable transformation \eqref{vt}, we have one-soliton solution
\begin{align}
 & A_{1}=\frac{{\rm e}^{i\eta_{1I}}}{2\sqrt{a_{1,1^{*}}}}{\rm sech}\left(\eta_{1R}+\frac{\chi_{1,1^{*}}}{2}\right),\\
 & A_{2}=re^{i\left(\varphi+\theta_{1I}\right)}\left[\cos\theta_{1I}+i\sin\theta_{1I}{\rm tanh}\left(\eta_{1R}+\frac{\chi_{1,1^{*}}}{2}\right)\right],\\
 & B_{0}=\frac{2(\mu+\lambda)k_{1R}^{2}}{\sigma c_{2}\left|k_{1}\right|^{2}}{\rm sech}^{2}\left(\eta_{1R}+\frac{\chi_{1,1^{*}}}{2}\right)-\frac{\lambda}{\sigma c_{2}},
\end{align}
with $\exp(\chi_{1,1^{*}})=a_{1,1^{*}}$ and $\eta_{1R}=\mathrm{Re}(\eta_1)$.

\begin{figure}
\centering{}\subfloat[]{\includegraphics[width=0.25\paperwidth]{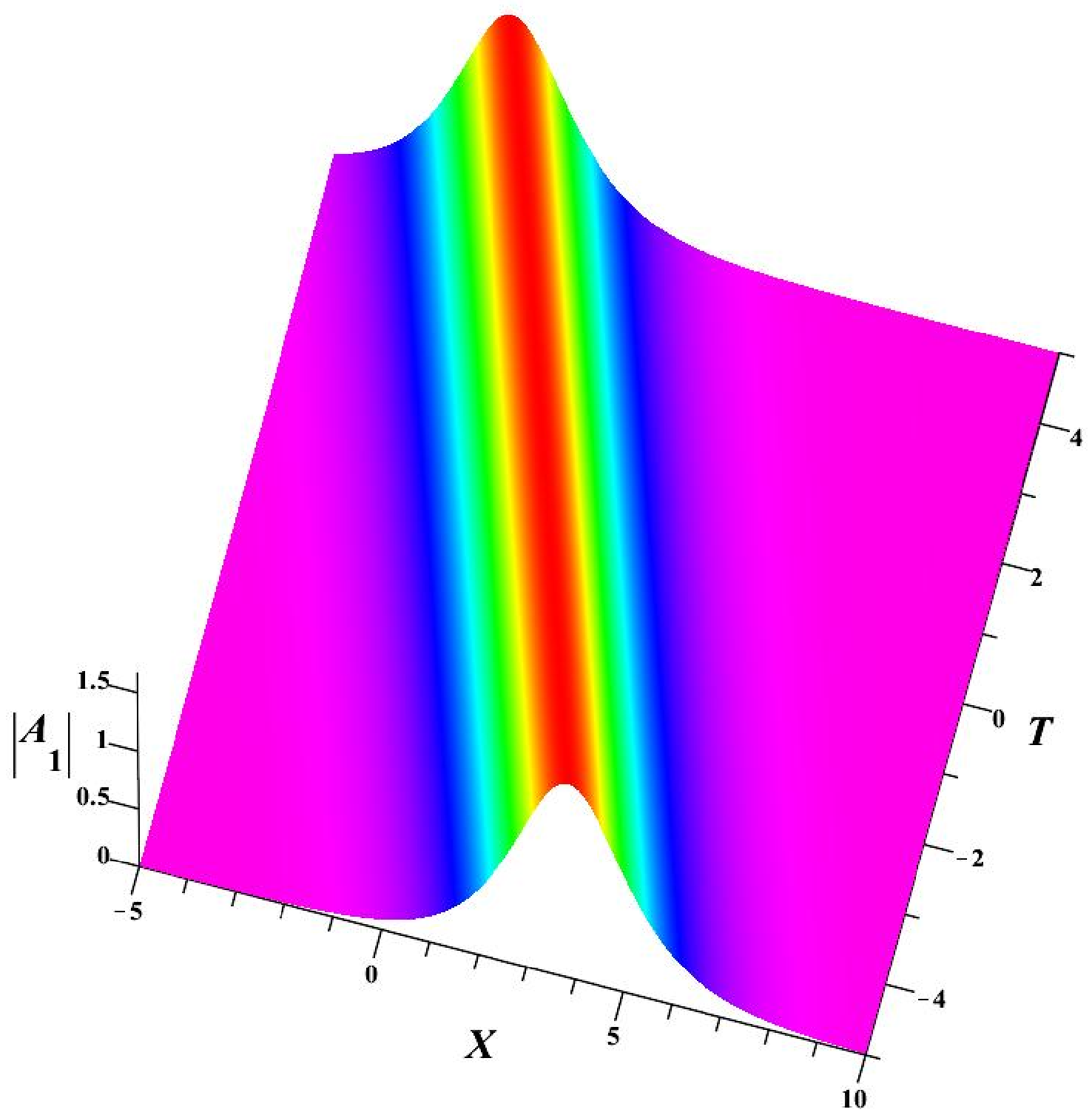}}
\subfloat[]{\includegraphics[width=0.25\paperwidth]{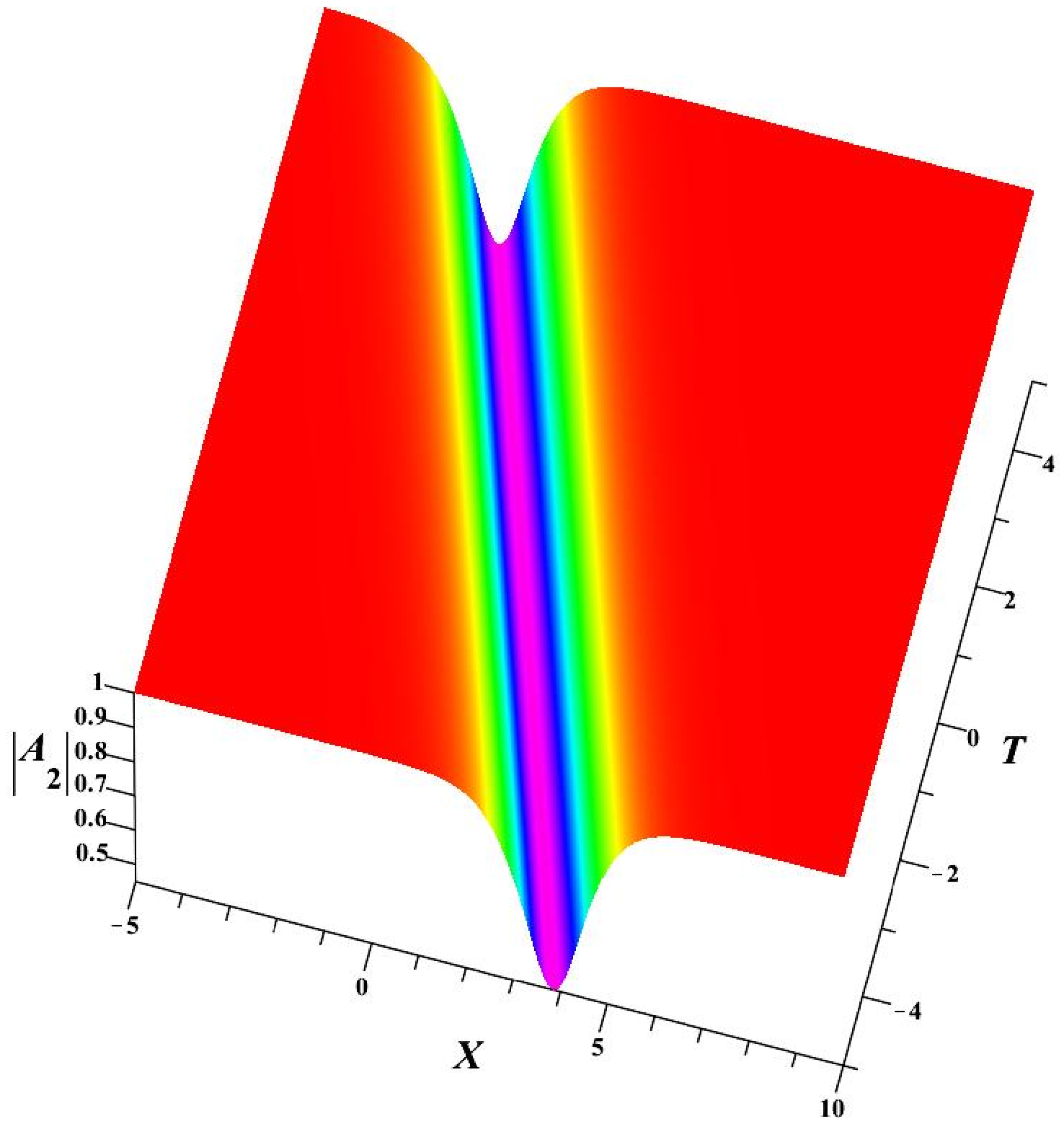}}
\subfloat[]{\includegraphics[width=0.25\paperwidth]{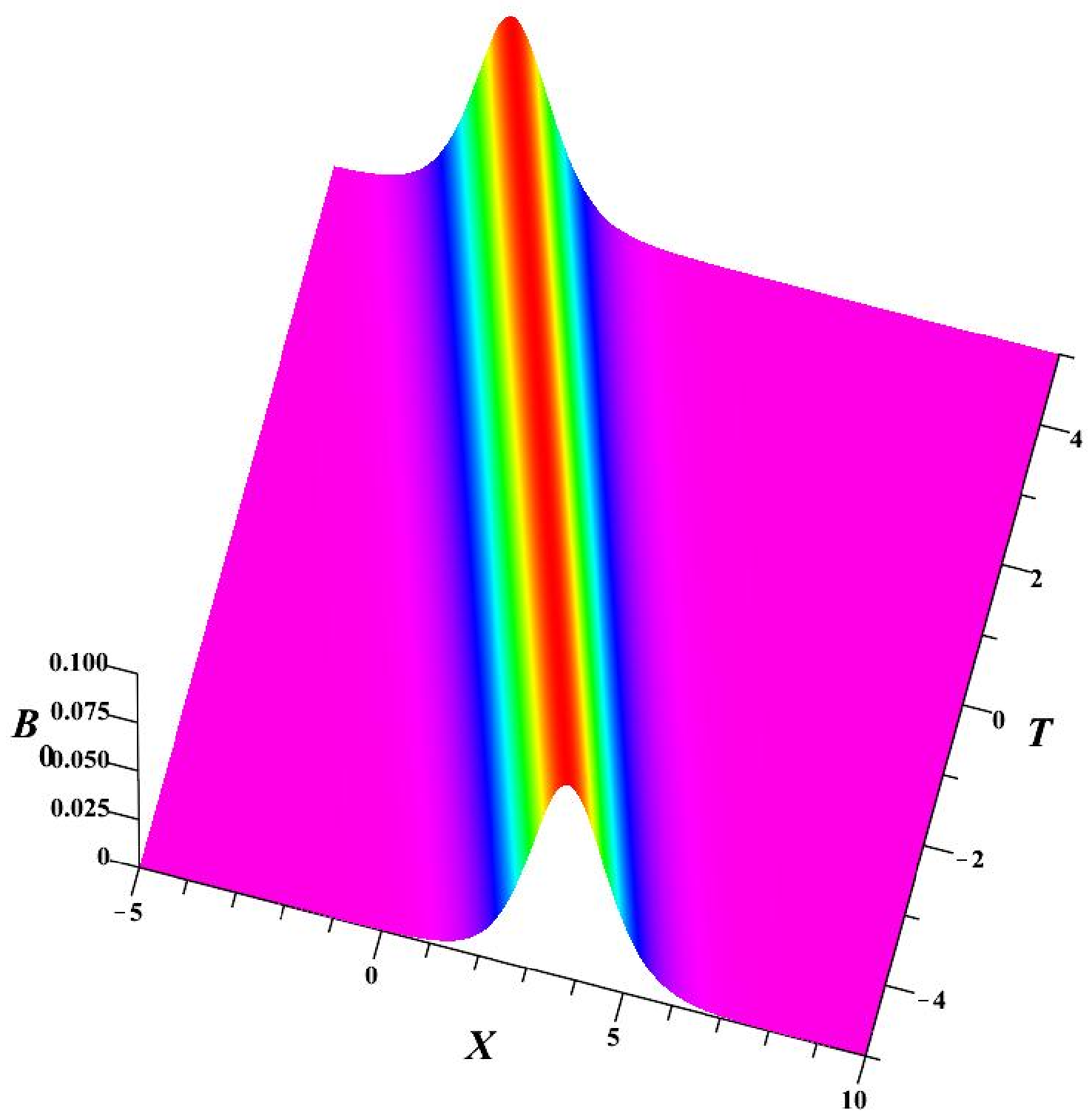}}
\caption{The evolution of one soliton solution to the focusing AB system
via $r=1, p=1,\thinspace k_{1}=1+i,\thinspace\lambda=0,\thinspace\varphi_{0}=0,\thinspace\eta_{1,0}=0,\, c_{1}=\frac{1}{100},c_2=10,\thinspace n_{1}=10$
and $n_{2}=100$.\label{fig:1sfc}}
\end{figure}

It's easy to see that $A_1$ and $B_0$ are bright soliton,  $A_2$ is dark soliton.  To get non-singular soliton solutions, conditions $\frac{2}{\sigma}+\frac{4p^{2}r^{2}}{\left|p^{2}+k_{1}^{2}\right|^{2}}>0$
and $k_{1R}\equiv\mathrm{Re}(k_1)\neq0$ must be satisfied to avoid vanishing of the
denominator. The travelling velocity of one-soliton is $\frac{\lambda+\mu}{\left|k_{1}\right|^{2}}$.
The amplitude of $A_{1}$is $\left|k_{1R}\right|\sqrt{\frac{2}{\sigma}+\frac{4p^{2}r^{2}}{\left|p^{2}+k_{1}^{2}\right|^{2}}}$.
The height of the trough of $A_{2}$ is $r\sqrt{1-\frac{4p^{2}k_{1R}^{2}}{\left|k_{1}^{2}+p^{2}\right|^{2}}}$.
The amplitude of $B_{0}$ is $\max\left|B_{0}+\frac{\lambda}{\sigma c_{2}}\right|=\frac{2|\mu+\lambda|k_{1R}^{2}}{|\sigma c_{2}|\left|k_{1}\right|^{2}}$. Fig.\ref{fig:1sfc} shows the evolution of  one-bright-dark soliton.

\subsection{Two-soliton solution}
To derive two-bright-dark-soliton solution, we suppose tau functions $g$ and $f$ have the following expression
\begin{align}
g_{1} & =e^{\eta_{1}}+e^{\eta_{2}}+a_{1,2,1^{*}}e^{\eta_{1}+\eta_{2}+\eta_{1}^{*}}+a_{1,2,2^{*}}e^{\eta_{1}+\eta_{2}+\eta_{2}^{*}},\label{eq:st2g1}\\
g_{2} & =re^{i\varphi}(1+a_{1,1^{*}}e^{\eta_{1}+\eta_{1}^{*}+\theta_{1}-\theta_{1}^{*}}+a_{1,2^{*}}e^{\eta_{1}+\eta_{2}^{*}+\theta_{1}-\theta_{2}^{*}}+a_{2,1^{*}}e^{\eta_{2}+\eta_{1}^{*}+\theta_{2}-\theta_{1}^{*}}\nonumber \\
 & \quad+a_{2,2^{*}}e^{\eta_{2}+\eta_{2}^{*}+\theta_{2}-\theta_{2}^{*}}+a_{1,2,1^{*},2^{*}}e^{\eta_{1}+\eta_{1}^{*}+\eta_{2}+\eta_{2}^{*}+\theta_{1}-\theta_{1}^{*}+\theta_{2}-\theta_{2}^{*}}),\label{eq:st2g2}\\
f & =1+a_{1,1^{*}}e^{\eta_{1}+\eta_{1}^{*}}+a_{1,2^{*}}e^{\eta_{1}+\eta_{2}^{*}}+a_{2,1^{*}}e^{\eta_{2}+\eta_{1}^{*}}+a_{2,2^{*}}e^{\eta_{2}+\eta_{2}^{*}}+a_{1,2,1^{*},2^{*}}e^{\eta_{1}+\eta_{1}^{*}+\eta_{2}+\eta_{2}^{*}}.\label{eq:st2f}
\end{align}
where
\begin{align}
 & \eta_{j}=k_{j}x+\frac{\mu+\lambda}{k_{j}}t+\eta_{j,0},\quad\thinspace\varphi=p\,x-\frac{\mu+\lambda}{p}\,t+\varphi_{0},\\
 & a_{j,l^{*}}=\left(k_{j}+k_{l}^{*}\right)^{-2}\left(\frac{2}{\sigma}+\frac{4p^{2}r^{2}}
 {\left(p^{2}+k_{j}^{2}\right)\left(p^{2}+k_{l}^{*2}\right)}\right)^{-1},\\
 & a_{j,l}=\left(k_{j}-k_{l}\right)^{2}\left(\frac{2}{\sigma}+\frac{4p^{2}r^{2}}{\left(p^{2}+k_{j}^{2}\right)\left(p^{2}+k_{l}^{2}\right)}\right),\\
 & a_{j^{*},l^{*}}=\left(k_{j}^{*}-k_{l}^{*}\right)^{2}\left(\frac{2}{\sigma}+\frac{4p^{2}r^{2}}{\left(p^{2}+k_{j}^{*2}\right)\left(p^{2}+k_{l}^{*2}\right)}\right),\\
 & a_{j_{1},j_{2},\cdots j_{n},l_{1}^{*},\cdots,l_{m}^{*}}=\prod_{1\leq m_{1}<m_{2}\leq n}a_{j_{m_{1}},j_{m_{2}}}\prod_{1\leq m_{1}<m_{2}\leq m}a_{l_{m_{1}}^{*},l_{m_{2}}^{*}}\prod_{1\leq m_{1}\leq n,\thinspace1\leq m_{2}\leq m}a_{j_{m_{1}},l_{m_{2}}^{*}},\\
 & \theta_{j}=\ln\frac{k_{j}-ip}{k_{j}+ip}=\ln\left|\frac{k_{j}-ip}{k_{j}+ip}\right|+i\arg\left(\frac{k_{j}-ip}{k_{j}+ip}\right).
\end{align}
Here we require the condition $\frac{2}{\sigma}+\frac{4p^{2}r^{2}}{\left|p^{2}+k_{j}^{2}\right|^{2}}>0$,
$k_{1}+k_{2}^{*}\neq0$ and $k_{jR}\neq0$ for $j=1,2$ such that
the obtained two-soliton solution is regular. We will show later that the two-soliton solutions exhibit the elastic
collision and soliton bound states.

\subsection{$N$-soliton solution}

In this subsection, we express the $N$-bright-dark soliton solution
to the two-component AB system (\ref{eq:eq21})-(\ref{eq:eq23}) by Pfaffian technique. A Pfaffian of
order $n$ is the square root of a determinant of order $2n$.
The definition and properties of Pfaffian could be found in \cite{Hirota}.
\begin{thm}\label{thm1}
The $N$-bright-dark soliton
solution to the two-component AB system(\ref{eq:eq21})-(\ref{eq:eq23}) is expressed by the Pfaffians
\begin{align}
f & =  \left(a_{1},\cdots,a_{2N},b_{2N},\cdots,b_{1}\right)=\left(\bullet\right),\label{pf1}\\
g_{1} & =  \left(d_{0},\beta,a_{1},\cdots,a_{2N},b_{2N},\cdots,b_{1}\right)=\left(d_{0},\beta,\bullet\right),\\
g_{2} & =  re^{i\varphi}\left(d_{0},\gamma,a_{1},\cdots,a_{2N},b_{2N},\cdots,b_{1}\right)=re^{i\varphi}\left(d_{0},\gamma,\bullet\right),\label{pf3}
\end{align}
where we take the notation $\left(\bullet\right)$ to represent $\left(a_{1},\cdots,a_{2N},b_{2N},\cdots,b_{1}\right)$ for simplicity. The Pfaffian entries are defined as
\begin{alignat*}{4}
 & \left(a_{j},a_{l}\right)=\frac{w_{l}-w_{j}}{w_{l}+w_{j}}{\rm e}^{\eta_{j}+\eta_{l}}, & \thinspace & \left(b_{j},\beta\right)=\frac{1+\mu_{j}}{2}, & \thinspace & \left(a_{j},\gamma\right)=\frac{2ip}{w_{j}+ip}e^{\eta_j}, & \thinspace & \left(d_{l},\beta_{k}\right)=0,\\
 & \left(b_{j},b_{l}\right)=\frac{1-\mu_{j}\mu_{l}}{2\left(w_{j}^{2}-w_{l}^{2}\right)\left(\frac{2}{\sigma}+\frac{4p^{2}r^{2}}{\left(p^{2}+w_{j}^{2}\right)\left(p^{2}+w_{l}^{2}\right)}\right)}, & \thinspace & \left(a_{j},\beta\right)=0, & \thinspace & \left(d_{l},a_{j}\right)=w_{j}^{l}{\rm e}^{\eta_{j}}, & \thinspace & \left(d_{l},d_{j}\right)=0,\\
 & \left(a_{j},b_{l}\right)=\delta_{jl}, & \thinspace & \left(b_{j},\gamma\right)=0, & \thinspace & \left(d_{l},b_{j}\right)=0, & \thinspace & \left(d_{l},\gamma\right)=\delta_{0l},
\end{alignat*}
and
\begin{gather*}
\delta_{jl}=\begin{cases}
0 & i\neq j\\
1 & i=j
\end{cases},\thinspace\quad\mu_{j}=\begin{cases}
1 & 1\leq j\leq N\\
-1 & N+1\leq j\leq2N
\end{cases},\thinspace\quad w_{j}=\begin{cases}
k_{j} & 1\leq j\leq N\\
k_{j}^{*} & N+1\leq j\leq2N
\end{cases}.
\end{gather*}
\end{thm}
Upon the properties of Pfaffian, we can express tau function and its derivatives as following
\begin{align*}
 & f_{x}=\left(d_{1},d_{0},\bullet\right),\qquad f_{t}=\left(\lambda+\mu\right)\left(d_{0},d_{-1},\bullet\right),\\
 & f_{xx}=\left(d_{2},d_{0},\bullet\right),\quad f_{xt}=\left(\lambda+\mu\right)\left(d_{1},d_{-1},\bullet\right),\\
 & g_{1,x}=\left(d_{1},\beta,\bullet\right),\qquad g_{1,t}=\left(\lambda+\mu\right)\left(d_{-1},\beta,\bullet\right),\\
 & g_{1,xt}=\left(\lambda+\mu\right)\left[\left(d_{0},\beta,\bullet\right)+\left(d_{1},d_{0},d_{-1},\beta,\bullet\right)\right],\\
 &\partial_{x} \left(d_{0},\gamma,\bullet\right)=ip\left(\bullet\right)-ip\left(d_{0},\gamma,\bullet\right)+\left(d_{1},d_{0},\bullet\right)+\left(d_{1},\gamma,\bullet\right),\\
 & \partial_{t} \left(d_{0},\gamma,\bullet\right)=\left(\lambda+\mu\right)\left[\frac{1}{ip}\left(\bullet\right)-\frac{1}{ip}\left(d_{0},\gamma,\bullet\right)-\left(d_{0},d_{-1},\bullet\right)+\left(d_{-1},\gamma,\bullet\right)\right],\\
 & \partial_{x}\partial_t \left(d_{0},\gamma,\bullet\right)=\left(\lambda+\mu\right)[2\left(d_{0},\gamma,\bullet\right)-2\left(\bullet\right)+2ip\left(d_{0},d_{-1},\bullet\right)\\
 & \qquad\qquad\qquad -ip\left(d_{-1},\gamma,\bullet\right)-\frac{1}{ip}\left(d_{1},\gamma,\bullet\right)+\left(d_{1},d_{0},d_{-1},\gamma,\bullet\right)].
\end{align*}

One can check that the substitution of above expressions into eq.(\ref{eq:b2c1}) leads to the
Pfaffian identity,
\begin{align}
\left(d_{1},d_{0},d_{-1},\beta,\bullet\right)\left(\bullet\right)
-\left(d_{1},\beta,\bullet\right)\left(d_{0},d_{-1},\bullet\right)-\left(d_{-1},\beta,\bullet\right)
\left(d_{1},d_{0},\bullet\right)+\left(d_{0},\beta,\bullet\right)\left(d_{1},d_{-1},\bullet\right)=0,
\end{align}
and eq.(\ref{eq:b2c1}) becomes
\begin{align}
\left(d_{1},d_{0},d_{-1},\gamma,\bullet\right)\left(\bullet\right)
-\left(d_{1},\gamma,\bullet\right)\left(d_{0},d_{-1},\bullet\right)-\left(d_{-1},\gamma,\bullet\right)\left(d_{1},d_{0},\bullet\right)
+\left(d_{0},\gamma,\bullet\right)\left(d_{1},d_{-1},\bullet\right)=0,
\end{align}
see \cite{XYZ} for reference. To prove Pfaffians \eqref{pf1}-\eqref{pf3} solve eq.(\ref{eq:b2c3}), we introduce new Pfaffian entries
\begin{alignat*}{3}
 & \left(b_{j},\beta^{*}\right)=\frac{1-\mu_{j}}{2}, & \thinspace & \left(a_{j},\beta^{*}\right)=0, & \thinspace & \left(d_{0},\beta^{*}\right)=0,\\
 & \left(a_{j},\gamma^{*}\right)=-\frac{2ip}{w_{j}-ip}, & \thinspace & \left(b_{j},\gamma^{*}\right)=0, & \thinspace & \left(d_{0},\gamma^{*}\right)=1,
\end{alignat*}
and express the conjugate of $g_1$ and $g_2$ as
\begin{gather*}
g_{1}^{*}=\left(d_{0},\beta^{*},\bullet\right),\thinspace g_{2}^{*}=re^{-i\varphi}\left(d_{0},\gamma^{*},\bullet\right).
\end{gather*}
The proof can be found in the Appendix.

\section{Collision of solitons to the two-component AB system}

\begin{figure}
\centering{}\subfloat[]{\includegraphics[width=0.25\paperwidth]{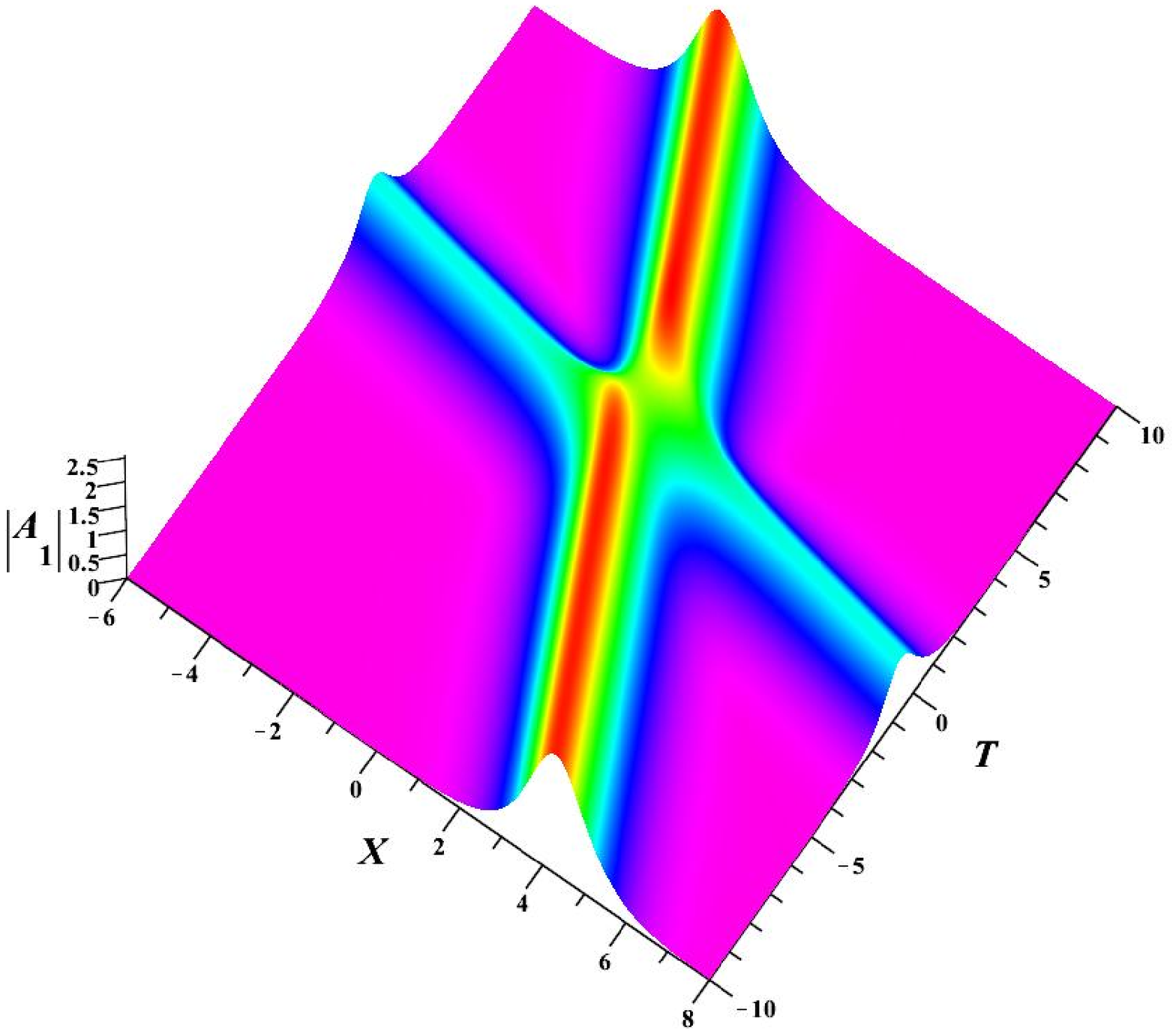}}\subfloat[]{\includegraphics[width=0.25\paperwidth]{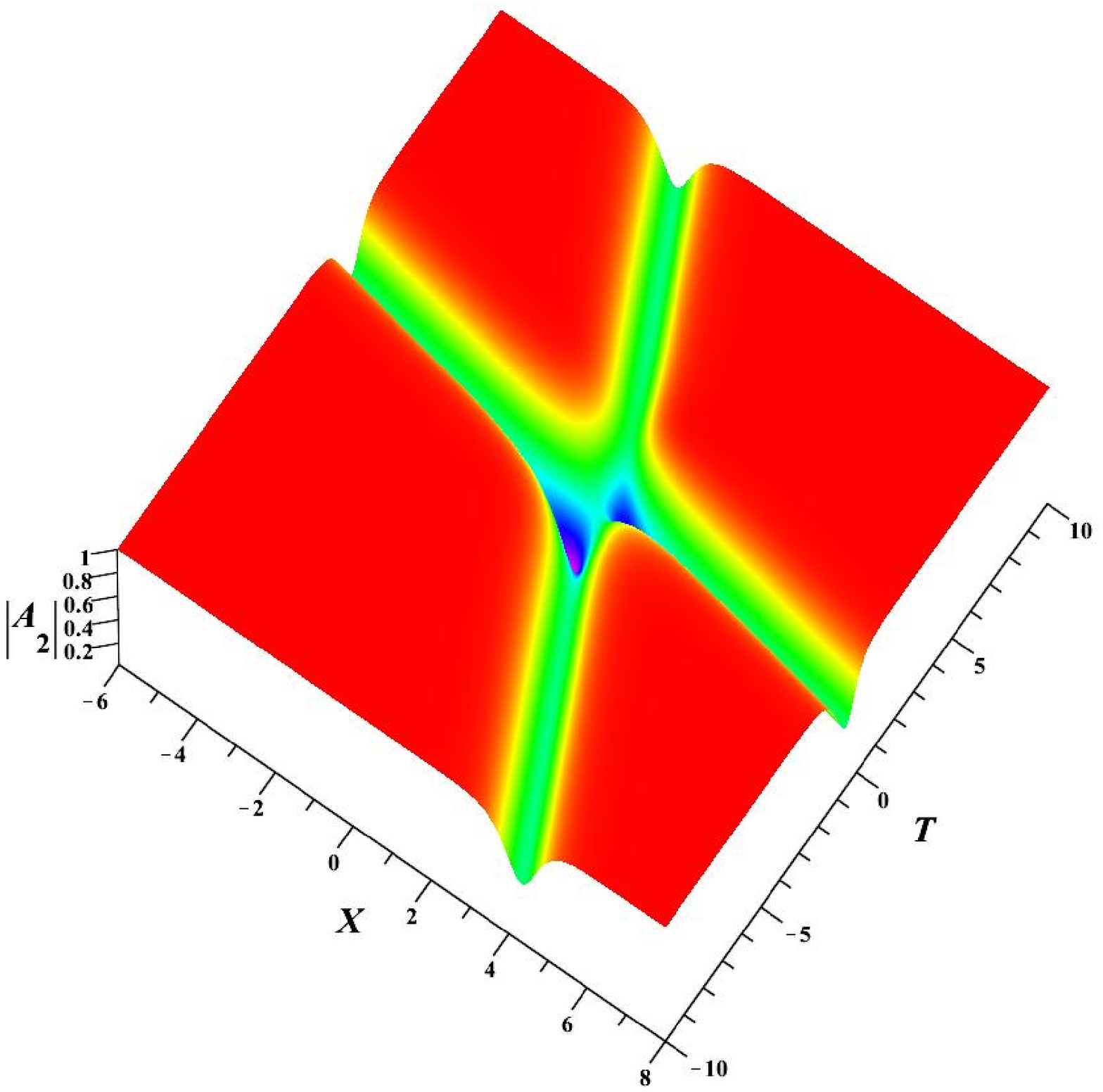}}\subfloat[]{\includegraphics[width=0.25\paperwidth]{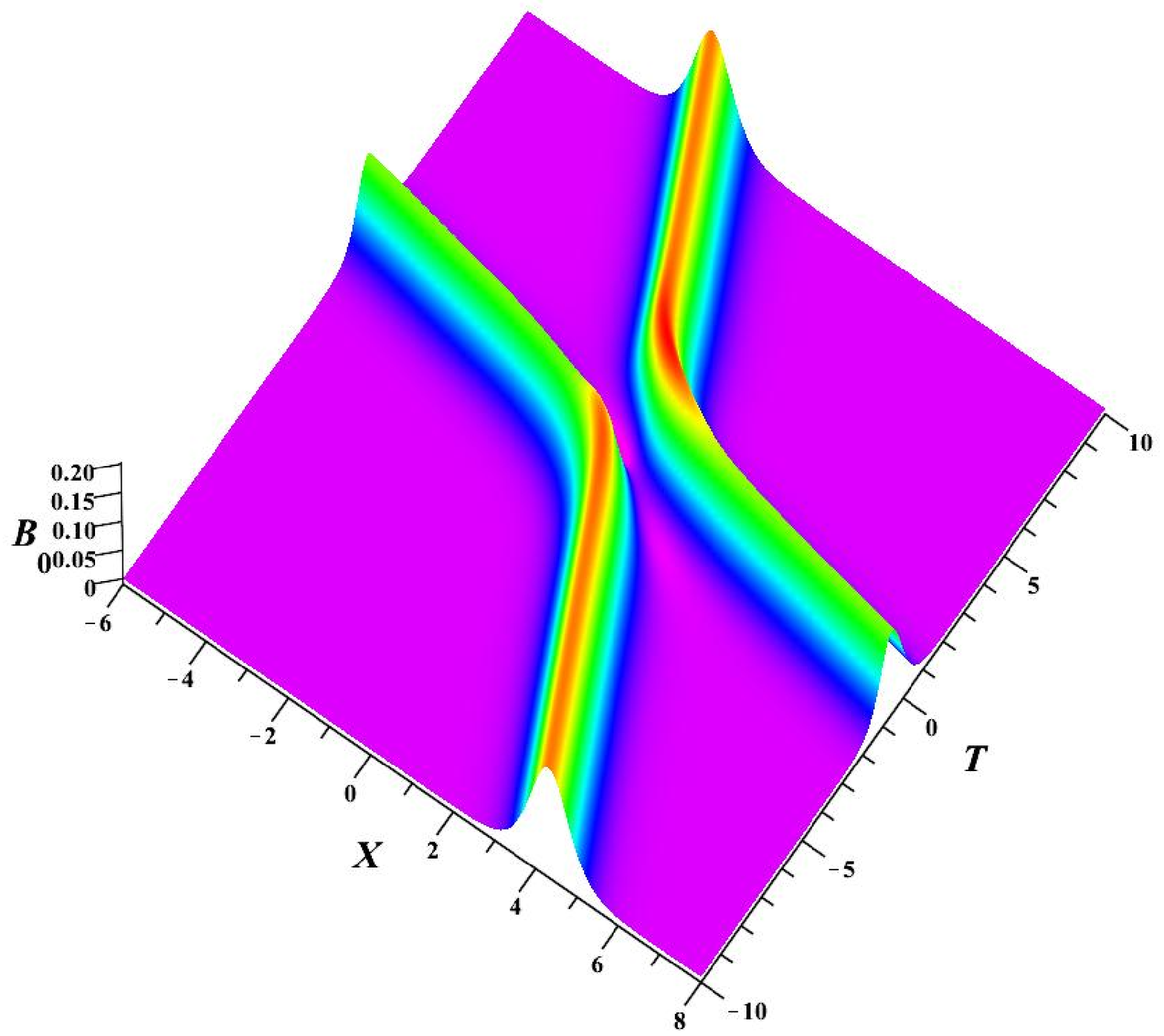}}\caption{The elastic collision of two soliton solution to the focusing AB system
via $r=1,\thinspace k_{1}=\frac{1}{2}+\frac{i}{3},\thinspace k_{2}=\frac{5}{3}+\frac{i}{2},\thinspace p=1,\thinspace\lambda=0,\thinspace\varphi_{0}=0,\thinspace\eta_{1,0}=0,\thinspace\eta_{2,0}=0,\:c_{1}=\frac{1}{100},\thinspace c_{2}=10,\thinspace n_{1}=10$
and $n_{2}=100$.\label{fig:2sfc}}
\end{figure}

\begin{figure}
\centering{}\subfloat[]{\includegraphics[width=0.25\paperwidth]{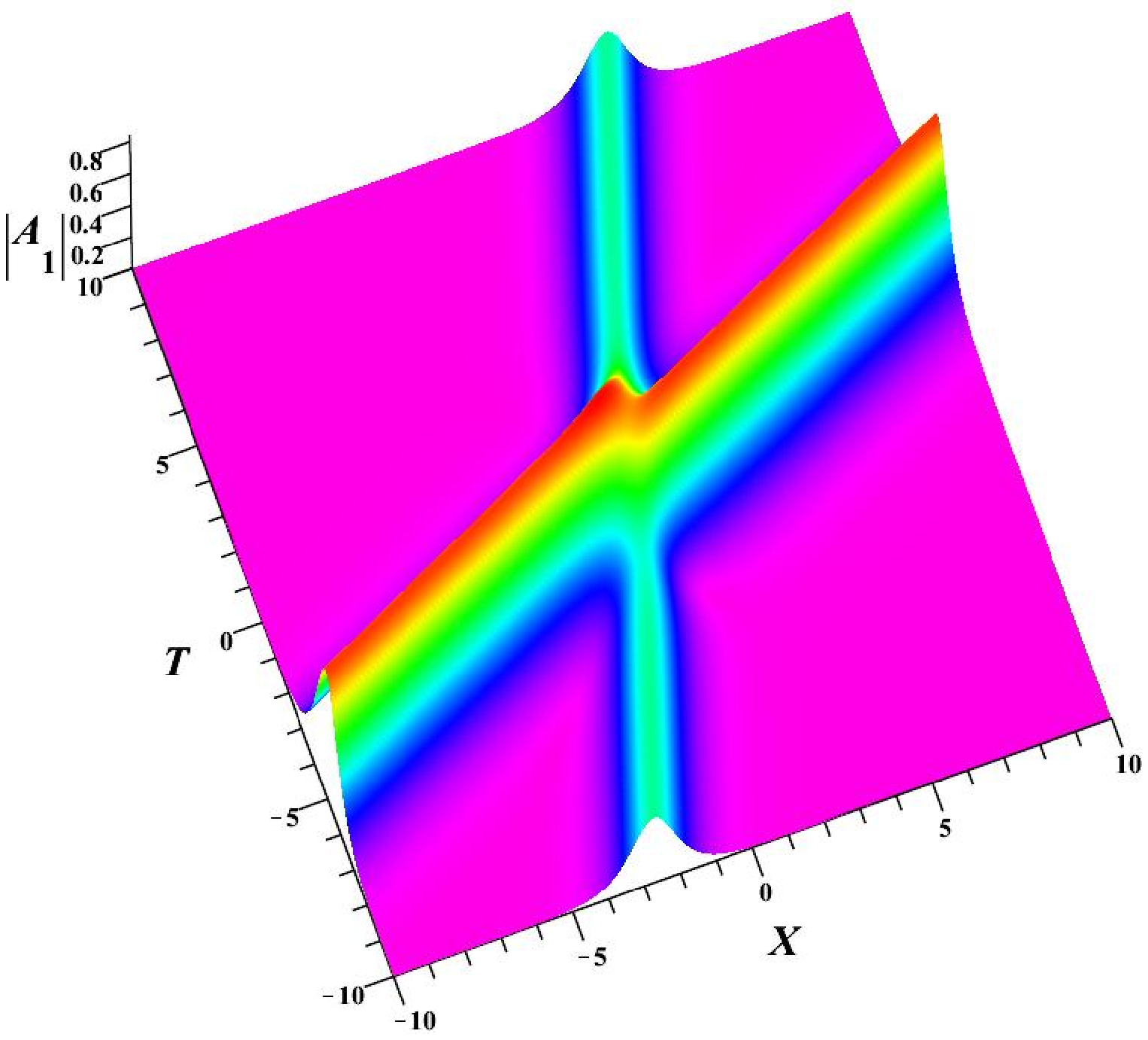}}\subfloat[]{\includegraphics[width=0.25\paperwidth]{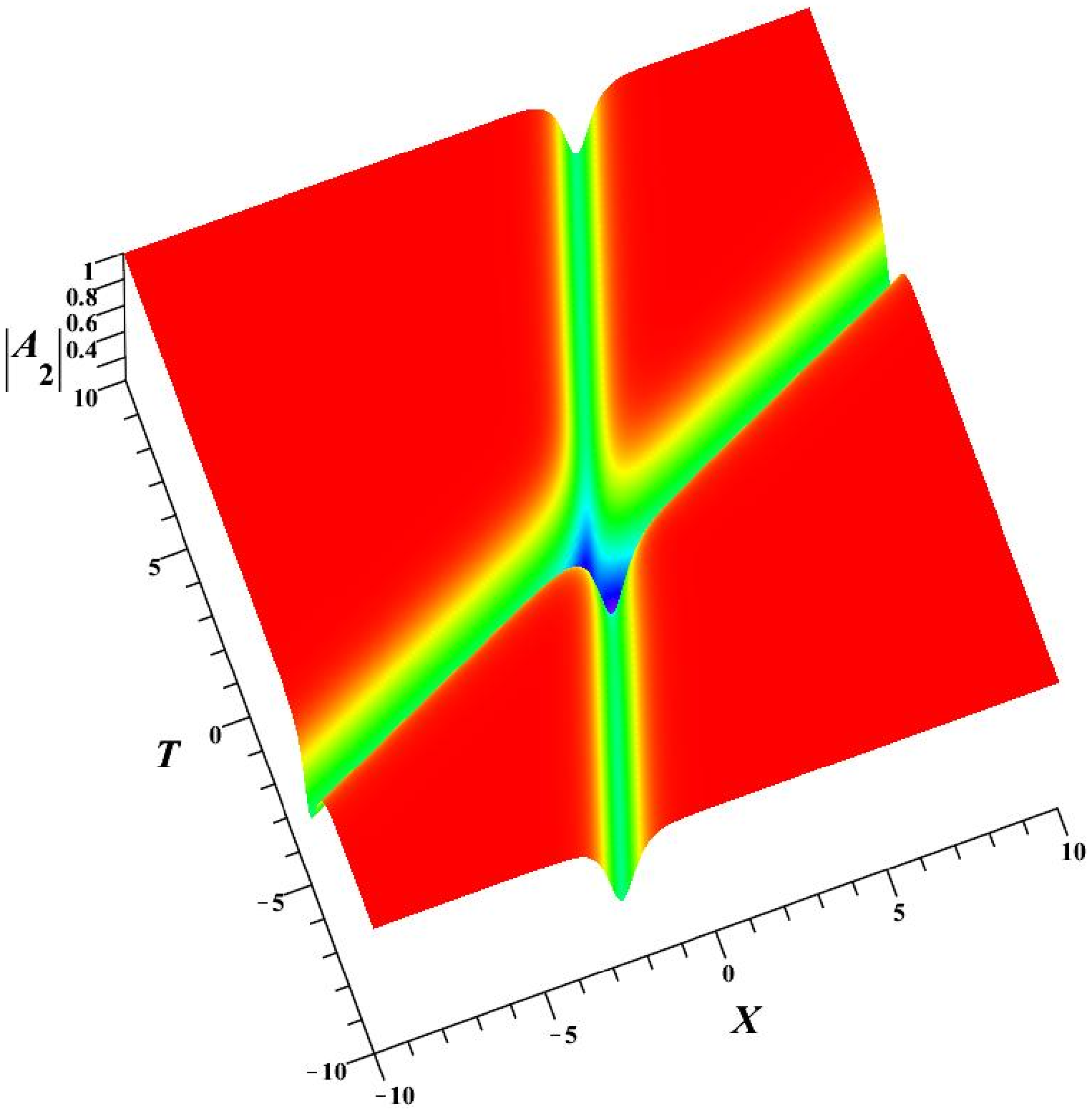}}\subfloat[]{\includegraphics[width=0.25\paperwidth]{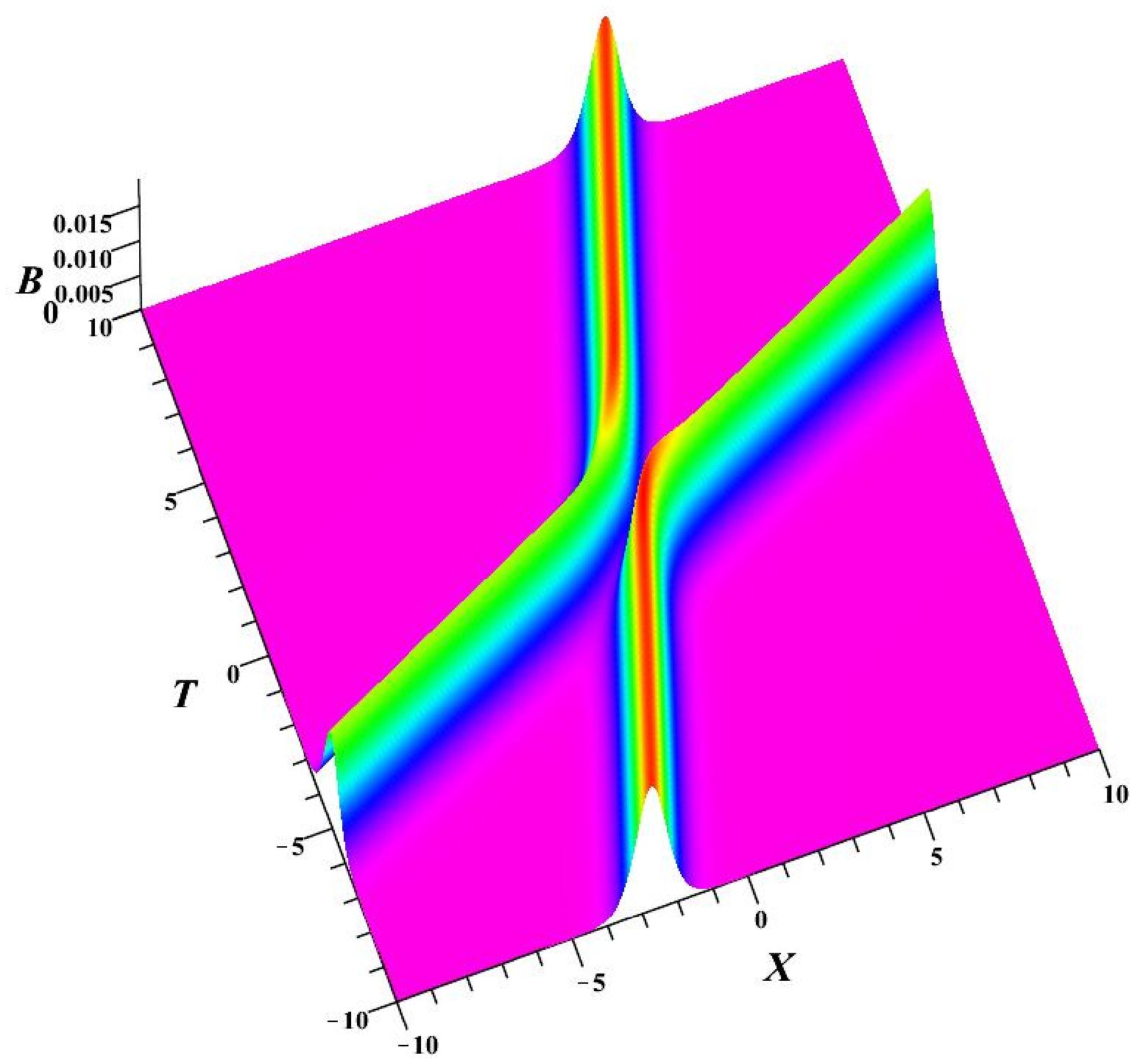}}\caption{The elastic collision of two soliton solution to the defocusing AB
system via $r=1,\thinspace k_{1}=\frac{1}{2}+\frac{i}{3},\thinspace k_{2}=\frac{5}{3}+\frac{i}{2},\thinspace p=1,\thinspace\lambda=0,\thinspace\varphi_{0}=0,\thinspace\eta_{1,0}=0,\thinspace\eta_{2,0}=0,\:c_{1}=\frac{1}{100},\thinspace c_{2}=10,\thinspace n_{1}=-10$
and $n_{2}=-1000$.\label{fig:2sdc}}
\end{figure}

For two-soliton solution (\ref{eq:st2g1})-(\ref{eq:st2f}) with
parameters $\left|k_{1}\right|\neq\left|k_{2}\right|$, the two soltions
travel with different velocities that leads to the collision between solitons. To prove the elastic collision and no energy change after collision, we check the asymptotic behavior of the two-soliton solutions. For two-soliton solution
\[
A_1=\frac{g_1}{f},\qquad A_2=\frac{g_2}{f},\qquad B_0=2(\ln f)_{xt}-\lambda,
\]
with $g_1,g_2$ and $f$ given by (\ref{eq:st2g1})-(\ref{eq:st2f}),  we have the following asymptotic
forms when $\frac{\lambda+\mu}{\left|k_{1}\right|^{2}}<\frac{\lambda+\mu}{\left|k_{2}\right|^{2}}$,

\begin{eqnarray}
A_{1} & \sim & \begin{cases}
\frac{a_{1,2}a_{1,2^{*}}e^{i\eta_{1I}}}{2\left|a_{1,2}a_{1,2^{*}}\right|\sqrt{a_{1,1^{*}}}}{\rm sech}\left(\eta_{1R}+\frac{\chi_{1,2,1^{*},2^{*}}-\chi_{2,2^{*}}}{2}\right) & \eta_{1R}\sim O(1),t\to+\infty\\
\frac{e^{i\eta_{1I}}}{2\sqrt{a_{1,1^{*}}}}{\rm sech}\left(\eta_{1R}+\frac{\chi_{1,1^{*}}}{2}\right) & \eta_{1R}\sim O(1),t\to-\infty\\
\frac{a_{1,2}a_{2,1^{*}}e^{i\eta_{1I}}}{2\left|a_{1,2}a_{1,2^{*}}\right|\sqrt{a_{2,2^{*}}}}{\rm sech}\left(\eta_{2R}+\frac{\chi_{1,2,1^{*},2^{*}}-\chi_{1,1^{*}}}{2}\right) & \eta_{2R}\sim O(1),t\to-\infty\\
\frac{e^{i\eta_{2I}}}{2\sqrt{a_{2,2^{*}}}}{\rm sech}\left(\eta_{2R}+\frac{\chi_{2,2^{*}}}{2}\right) & \eta_{2R}\sim O(1),t\to+\infty
\end{cases}\\
A_{2} & \sim & \begin{cases}
re^{i\left(\varphi+2\theta_{2I}+\theta_{1I}\right)}\left[\cos\theta_{1I}+i\sin\theta_{1I}{\rm tanh}\left(\eta_{1R}+\frac{\chi_{1,2,1^{*},2^{*}}-\chi_{2,2^{*}}}{2}\right)\right] & \eta_{1R}\sim O(1),t\to+\infty\\
re^{i\left(\varphi+\theta_{1I}\right)}\left[\cos\theta_{1I}+i\sin\theta_{1I}{\rm tanh}\left(\eta_{1R}+\frac{\chi_{1,1^{*}}}{2}\right)\right] & \eta_{1R}\sim O(1),t\to-\infty\\
re^{i\left(\varphi+2\theta_{1I}+\theta_{2I}\right)}\left[\cos\theta_{2I}+i\sin\theta_{2I}{\rm tanh}\left(\eta_{2R}+\frac{\chi_{1,2,1^{*},2^{*}}-\chi_{1,1^{*}}}{2}\right)\right] & \eta_{2R}\sim O(1),t\to-\infty\\
re^{i\left(\varphi+\theta_{2I}\right)}\left[\cos\theta_{2I}+i\sin\theta_{2I}{\rm tanh}\left(\eta_{2R}+\frac{\chi_{2,2^{*}}}{2}\right)\right] & \eta_{2R}\sim O(1),t\to+\infty
\end{cases}\\
B_{0} & \sim & \begin{cases}
\frac{2(\mu+\lambda)k_{1R}^{2}}{\sigma c_{2}\left|k_{1}\right|^{2}}{\rm sech}^{2}\left(\eta_{1R}+\frac{\chi_{1,2,1^{*},2^{*}}-\chi_{2,2^{*}}}{2}\right)-\frac{\lambda}{\sigma c_{2}} & \eta_{1R}\sim O(1),t\to+\infty\\
\frac{2(\mu+\lambda)k_{1R}^{2}}{\sigma c_{2}\left|k_{1}\right|^{2}}{\rm sech}^{2}\left(\eta_{1R}+\frac{\chi_{1,1^{*}}}{2}\right)-\frac{\lambda}{\sigma c_{2}} & \eta_{1R}\sim O(1),t\to-\infty\\
\frac{2(\mu+\lambda)k_{2R}^{2}}{\sigma c_{2}\left|k_{2}\right|^{2}}{\rm sech}^{2}\left(\eta_{2R}+\frac{\chi_{1,2,1^{*},2^{*}}-\chi_{1,1^{*}}}{2}\right)-\frac{\lambda}{\sigma c_{2}} & \eta_{2R}\sim O(1),t\to-\infty\\
\frac{2(\mu+\lambda)k_{2R}^{2}}{\sigma c_{2}\left|k_{2}\right|^{2}}{\rm sech}^{2}\left(\eta_{2R}+\frac{\chi_{2,2^{*}}}{2}\right)-\frac{\lambda}{\sigma c_{2}} & \eta_{2R}\sim O(1),t\to+\infty
\end{cases}
\end{eqnarray}
with $\exp(\chi_{i,j^{*}})=a_{i,j^{*}},\,
\exp(\chi_{1,2,1^{*},2^{*}})=a_{1,2,1^{*},2^{*}}$.
One can check that the collision is completely elastic for $A_{1}$, $A_{2}$ and $B_{0}$.
Both the velocity and the amplitude keep invariant after collision.
Fig.\ref{fig:2sfc} shows the collision of two-bright-dark-soliton
solution in the focusing case $(\sigma>0)$. Fig.\ref{fig:2sdc} depicts the
collision in the defocusing case $(\sigma<0)$.


\section{Soliton bound states to the two-component AB system}

For two-soliton solution (\ref{eq:st2g1})-(\ref{eq:st2f}) with
parameters $\left|k_{1}\right|=\left|k_{2}\right|$, the two solitons
have the same velocity and exhibit periodic phenomenon. We find that
the solution has the following periodic property
\begin{alignat*}{2}
 & f(x,t) &  & =f\left(x-\frac{\pi}{k_{2I}-k_{1I}},t+\frac{\pi\left|k_{1}\right|^{2}}{\left(\lambda+\mu\right)\left(k_{2I}-k_{1I}\right)}\right),\\
 & \left|g_{1}\left(x,t\right)\right| &  & =\left|g_{1}\left(x-\frac{\pi}{k_{2I}-k_{1I}},t+\frac{\pi\left|k_{1}\right|^{2}}{\left(\lambda+\mu\right)\left(k_{2I}-k_{1I}\right)}\right)\right|,\\
 & \left|g_{2}\left(x,t\right)\right| &  & =\left|g_{2}\left(x-\frac{\pi}{k_{2I}-k_{1I}},t+\frac{\pi\left|k_{1}\right|^{2}}{\left(\lambda+\mu\right)\left(k_{2I}-k_{1I}\right)}\right)\right|.
\end{alignat*}
Thus we have
\begin{alignat*}{2}
 & B_{0}\left(X,T\right) &  & =B_{0}\left(X+c_{2}\pi\frac{c_{1}\left|k_{1}\right|^{2}-\left(\lambda+\mu\right)}{\left(\lambda+\mu\right)\left(c_{2}-c_{1}\right)\left(k_{2I}-k_{1I}\right)},T+\pi\frac{c_{2}\left|k_{1}\right|^{2}-\left(\lambda+\mu\right)}{\left(\lambda+\mu\right)\left(c_{2}-c_{1}\right)\left(k_{2I}-k_{1I}\right)}\right)\\
 & \left|A_{1}\left(X,T\right)\right| &  & =\left|A_{1}\left(X+c_{2}\pi\frac{c_{1}\left|k_{1}\right|^{2}-\left(\lambda+\mu\right)}{\left(\lambda+\mu\right)\left(c_{2}-c_{1}\right)\left(k_{2I}-k_{1I}\right)},T+\pi\frac{c_{2}\left|k_{1}\right|^{2}-\left(\lambda+\mu\right)}{\left(\lambda+\mu\right)\left(c_{2}-c_{1}\right)\left(k_{2I}-k_{1I}\right)}\right)\right|\\
 & \left|A_{2}\left(X,T\right)\right| &  & =\left|A_{2}\left(X+c_{2}\pi\frac{c_{1}\left|k_{1}\right|^{2}-\left(\lambda+\mu\right)}{\left(\lambda+\mu\right)\left(c_{2}-c_{1}\right)\left(k_{2I}-k_{1I}\right)},T+\pi\frac{c_{2}\left|k_{1}\right|^{2}-\left(\lambda+\mu\right)}{\left(\lambda+\mu\right)\left(c_{2}-c_{1}\right)\left(k_{2I}-k_{1I}\right)}\right)\right|
\end{alignat*}

\begin{figure}
\centering{}\subfloat[]{\includegraphics[width=0.26\paperwidth]{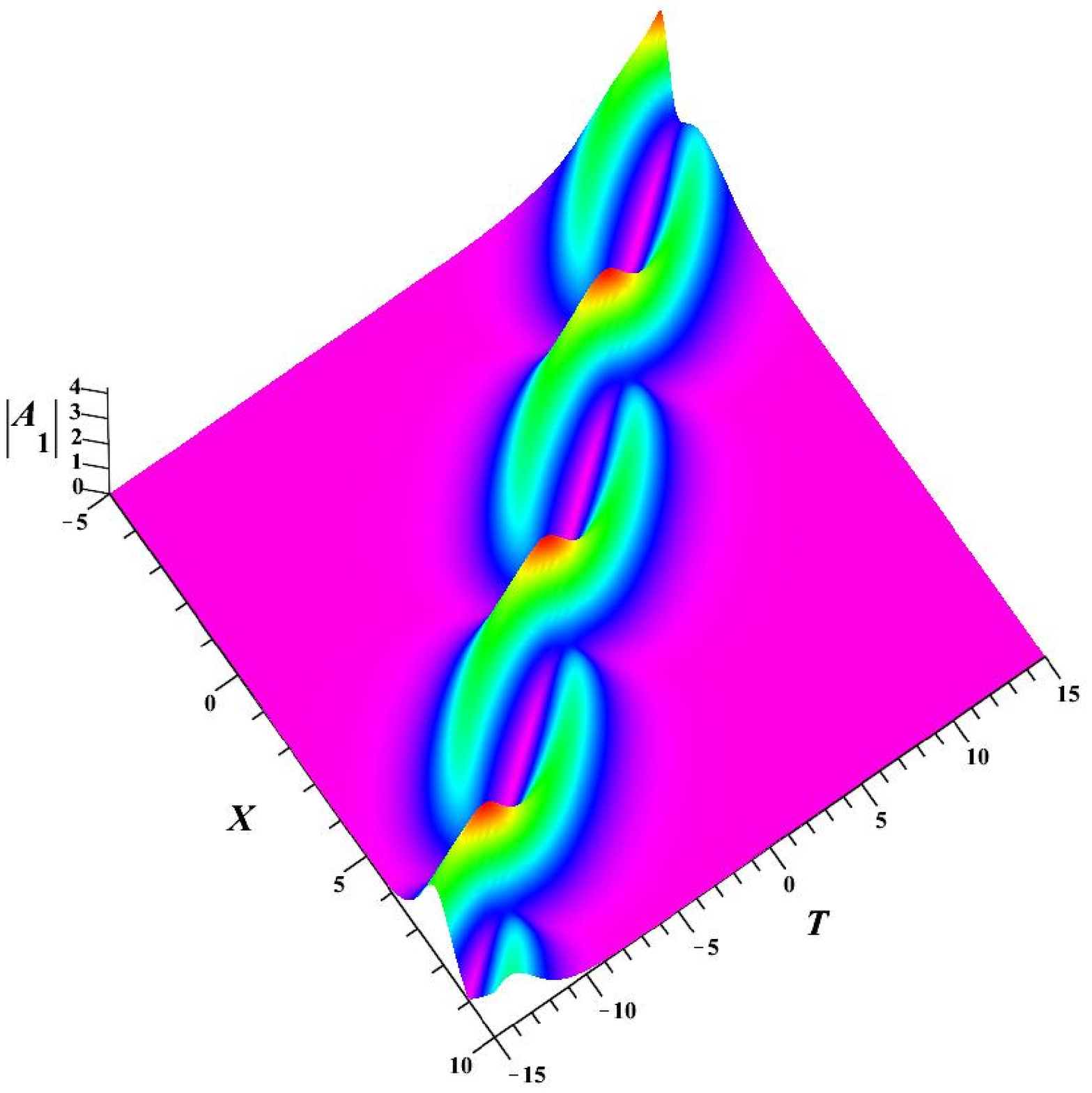}}\subfloat[]{\includegraphics[width=0.26\paperwidth]{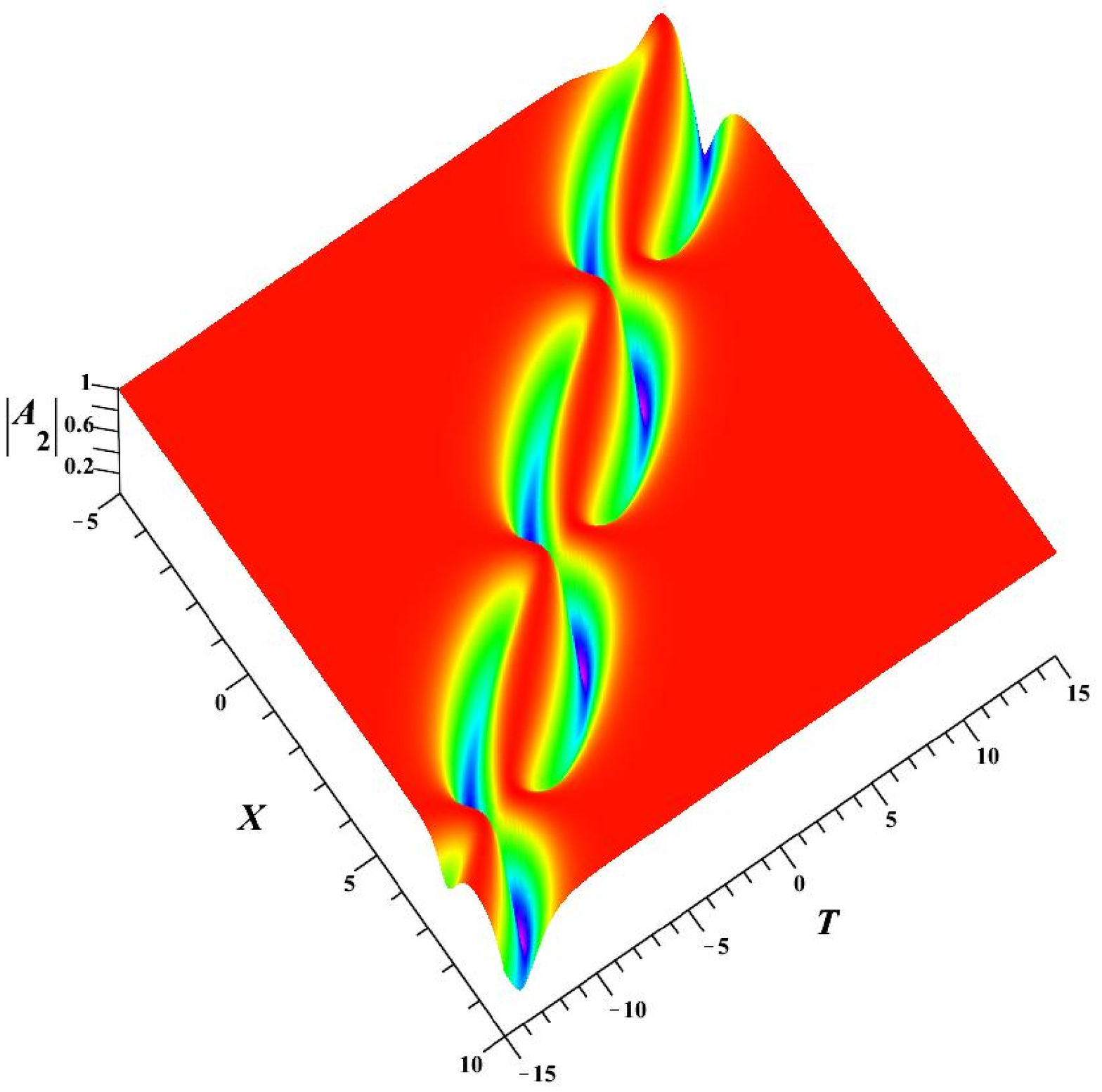}}\subfloat[]{\includegraphics[width=0.26\paperwidth]{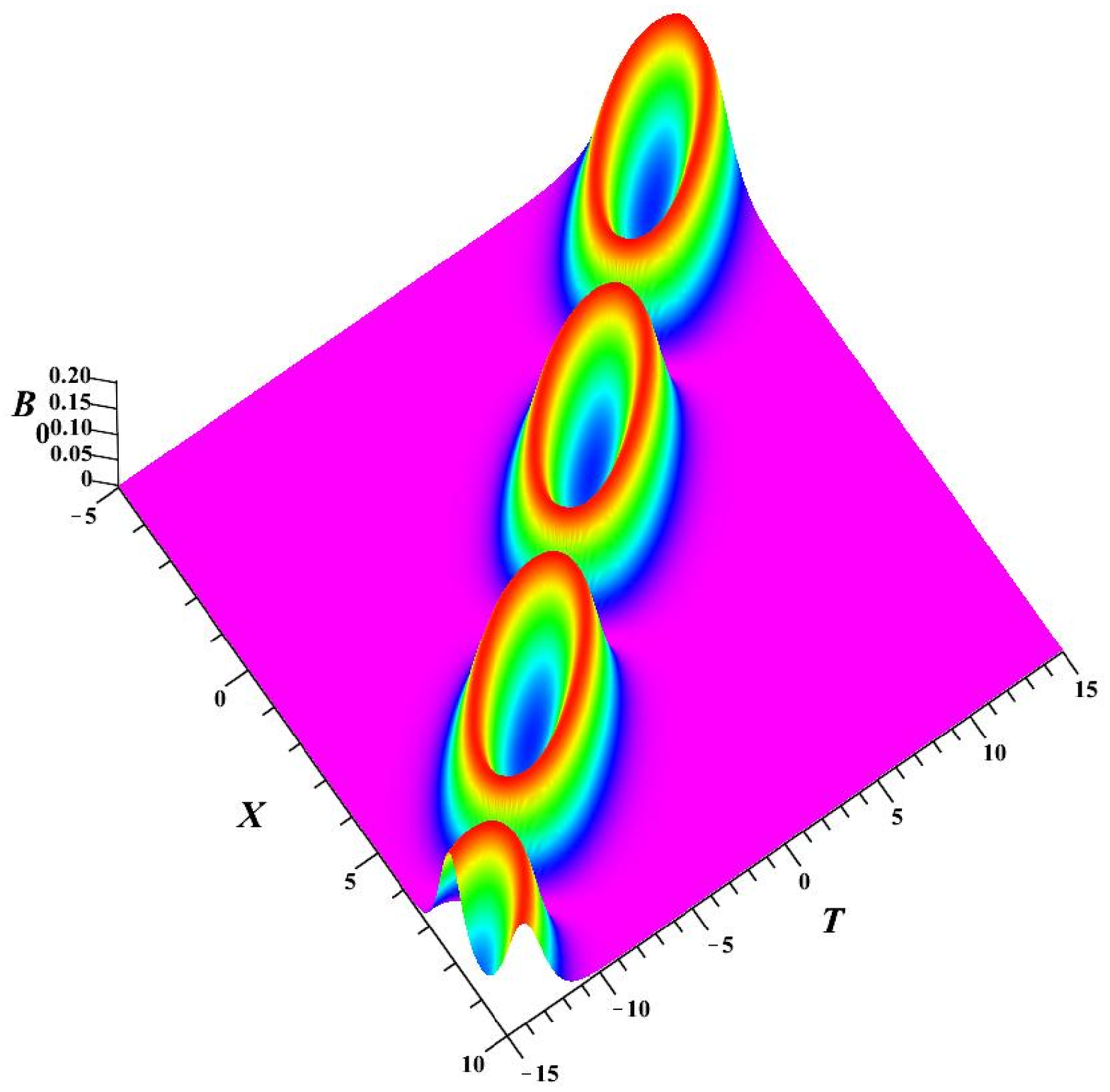}}
\caption{Bound states of solitons to the focusing AB system via
$r=1,\thinspace k_{1}=\frac{3}{2}\exp\left(\frac{i}{4}\right),\thinspace k_{2}=\frac{3}{2}\exp\left(-\frac{i}{4}\right),\thinspace p=1,\thinspace\lambda=0,\thinspace\varphi_{0}=0,\thinspace\eta_{1,0}=\frac{3}{20},\thinspace\eta_{2,0}=-\frac{1}{10},\:c_{1}=\frac{1}{100},\thinspace c_{2}=10,\thinspace n_{1}=10$
and $n_{2}=100$.\label{fig:2sfp}}
\end{figure}

\begin{figure}
\centering{}\subfloat[]{\includegraphics[width=0.26\paperwidth]{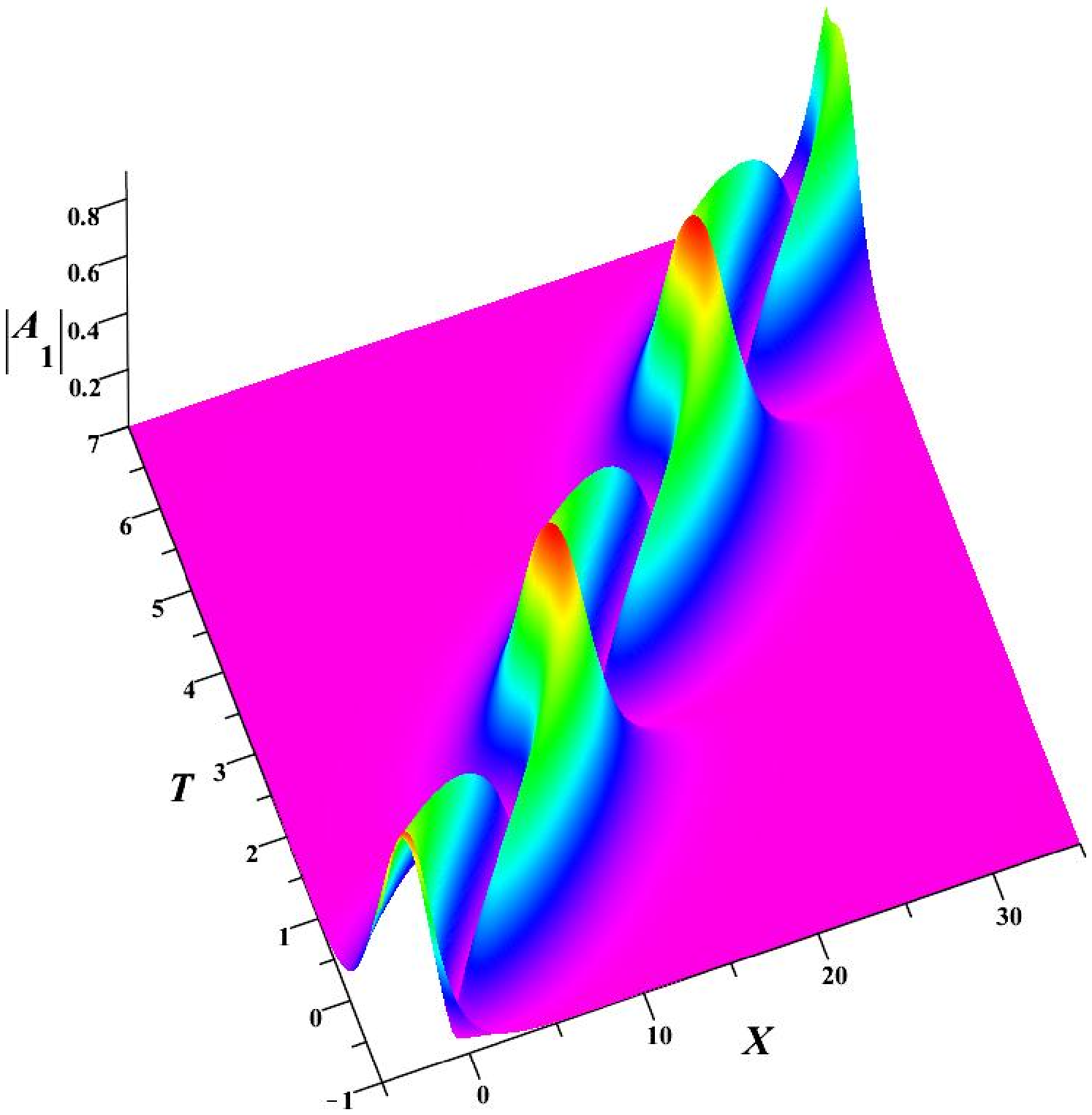}}\subfloat[]{\includegraphics[width=0.26\paperwidth]{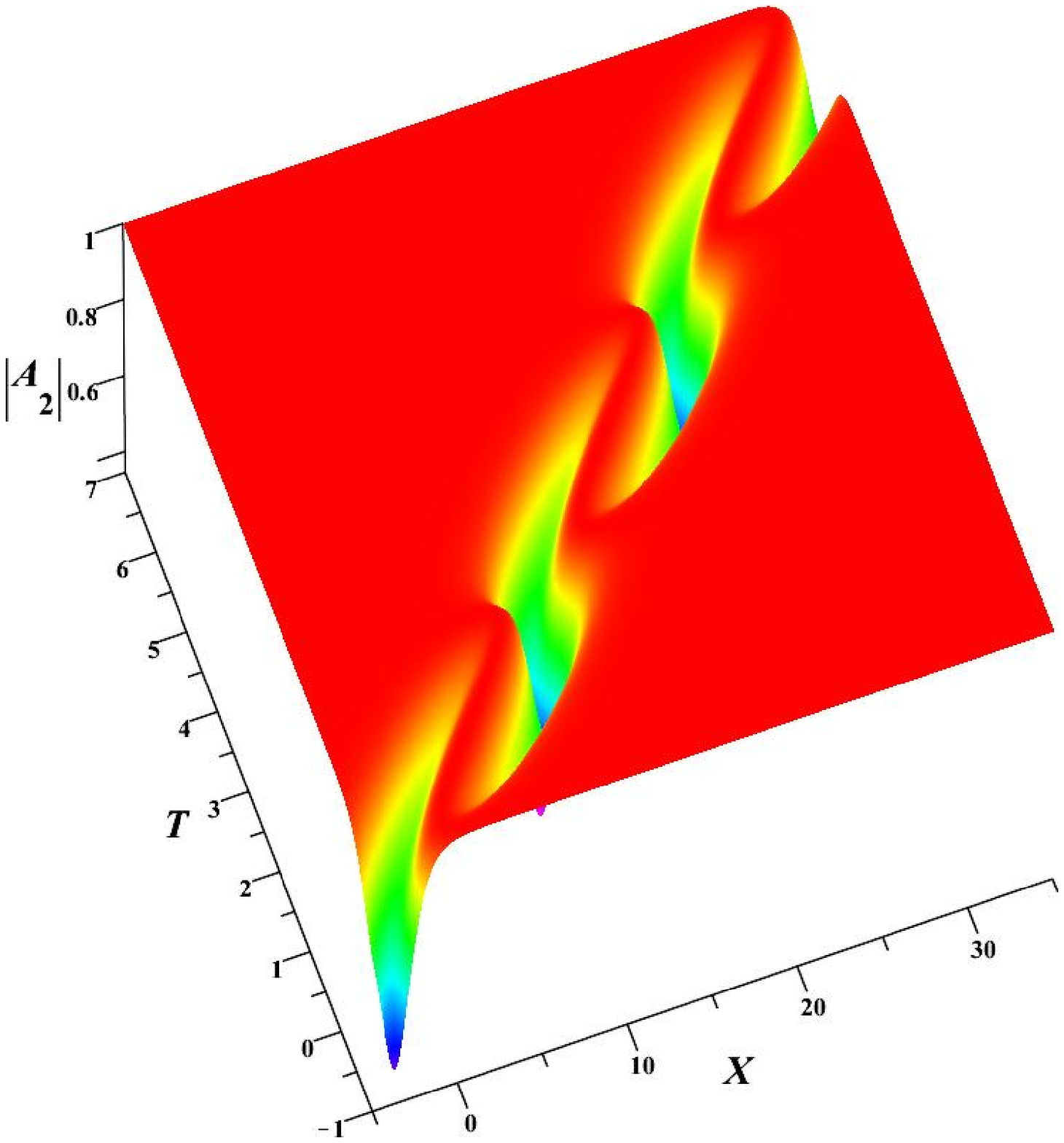}}\subfloat[]{\includegraphics[width=0.26\paperwidth]{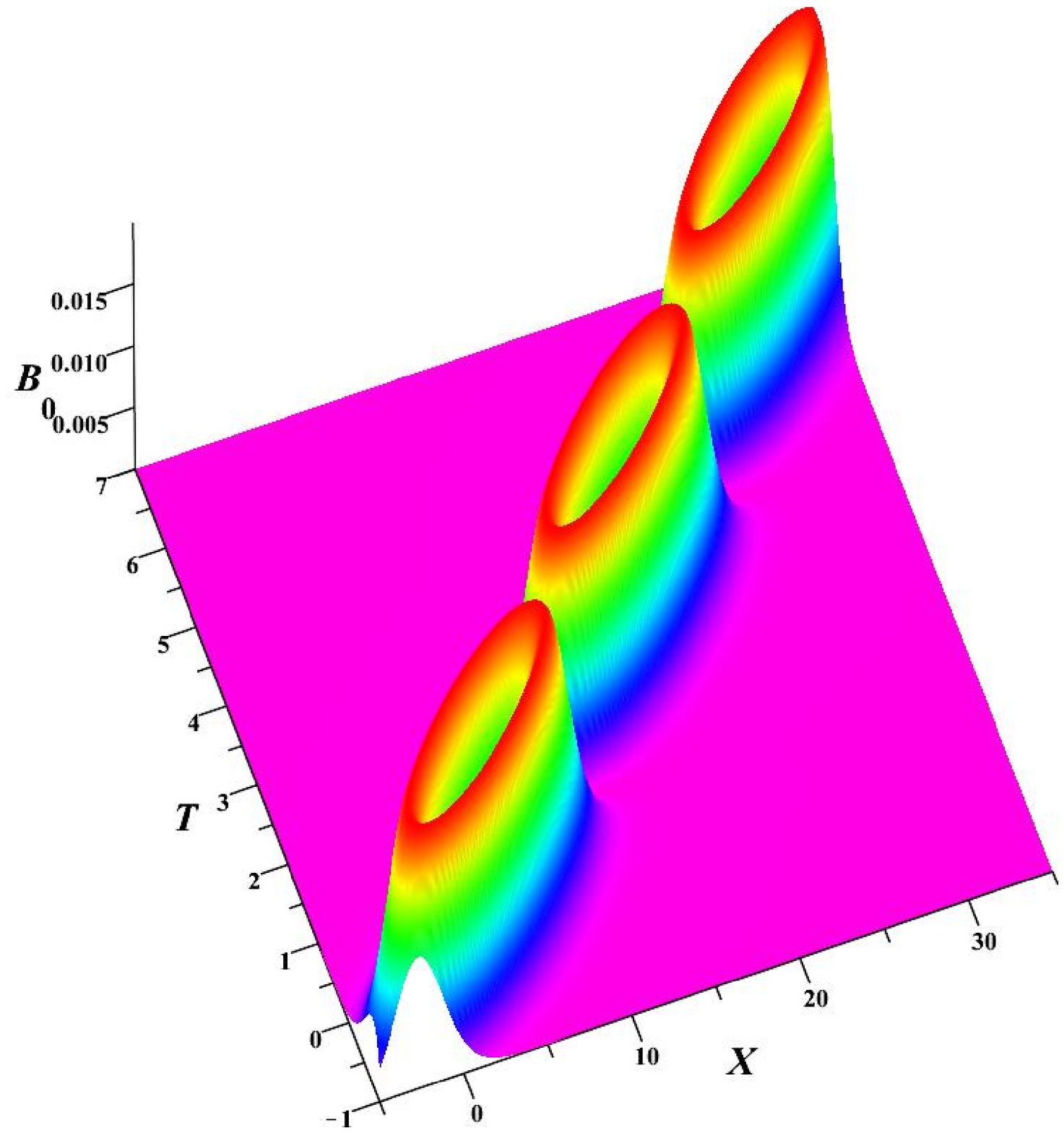}}\caption{ Bound states of solitons to the defocusing AB system via
$r=1,\thinspace k_{1}=\exp\left(-\frac{2i}{5}\right),\thinspace k_{2}=\frac{3}{2}\exp\left(-\frac{i}{3}\right),\thinspace p=1,\thinspace\lambda=0,\thinspace\varphi_{0}=0,\thinspace\eta_{1,0}=0,\thinspace\eta_{2,0}=0,\:c_{1}=\frac{1}{100},\thinspace c_{2}=10,\thinspace n_{1}=-10$
and $n_{2}=-1000$.\label{fig:2sdp}}
\end{figure}

Fig.\ref{fig:2sfp} displays the bound states of two-bright-dark soliton
to the focusing AB system. Fig.\ref{fig:2sdp} shows the
bound states of two-bright-dark solitons to the defocusing
AB system. When periods are small enough, bound states of solitons  appear parallel.

\section{Bright-dark $N$-soliton solution to the $M$-component AB system}

Ins this section, we give the bright-dark $N$-solition solution to
the $M$-component AB system (\ref{eq:eqm1})-(\ref{eq:eqm2}).
We first give the bilinear form to the system. Through the dependent
variable transformation
\begin{equation}
A_{j}=\frac{g_{j}}{f},\qquad B=2\left(\ln f\right)_{xt}-\lambda,\qquad j=1,\cdots,M,
\end{equation}
the $M$-component AB system \eqref{eq:mc1} -\eqref{eq:mc2}
is transformed into  bilinear form
\begin{align}
 & D_{x}D_{t}g_{j}\cdot f=\left(\mu+\lambda\right)g_{j}f,\qquad j=1,\cdots,M\label{eq:bim1}\\
 & \left(D_{x}^{2}+r^{2}\sigma\right)f\cdot f=\sigma\sum_{j=1}^{M}\left|g_{j}\right|^{2}.\label{eq:bim2}
\end{align}

We construct $N$-bright-dark-soliton solution with the first $M_1$ ($1\leq M_{1}\leq M$) components  $g_{1},\cdots,g_{M_{1}}$  bright and the left $g_{M_{1}+1},\cdots,g_{M}$ dark.
\begin{thm}\label{thm2}
The Pfaffians
\begin{align}
f & =  \left(a_{1},\cdots,a_{2N},b_{2N},\cdots,b_{1}\right)=\left(\bullet\right),\label{mn-1}\\
g_{j} & =  \left(d_{0},\beta_{j},a_{1},\cdots,a_{2N},b_{2N},\cdots,b_{1}\right)=\left(d_{0},\beta_{j},\bullet\right),\thinspace j=1,\cdots,M_{1},\label{mn-2}\\
g_{j} & =  e^{i\varphi_{j}}\left(d_{0},\gamma_{j},a_{1},\cdots,a_{2N},b_{2N},\cdots,b_{1}\right)=e^{i\varphi_{j}}\left(d_{0},\gamma_{j},\bullet\right),\thinspace j=M_{1}+1,\cdots,M,\label{mn-3}
\end{align}
solve the bilinear Eqs. \eqref{eq:bim1} and \eqref{eq:bim2} provided that the elements of the Pfaffians are defined by
\begin{alignat*}{3}
 & \left(a_{j},a_{l}\right)=\frac{w_{l}-w_{j}}{w_{l}+w_{j}}{\rm e}^{\eta_{j}+\eta_{l}}, & \thinspace & \left(a_{j},b_{l}\right)=\delta_{jl}, & \thinspace & \left(d_{l},d_{j}\right)=0,\\
 & \left(b_{j},b_{l}\right)=\frac{\left(1-\mu_{j}\mu_{l}\right)\sum\limits _{n=1}^{M_{1}}\alpha_{j}^{\left(n\right)}\alpha_{l}^{\left(n\right)}}{2\left(w_{j}^{2}-w_{l}^{2}\right)\left(\frac{2}{\sigma}+\sum\limits _{n=M_{1}+1}^{M}\frac{4p_{n}^{2}r_{n}^{2}}{\left(p_{n}^{2}+w_{j}^{2}\right)\left(p_{n}^{2}+w_{l}^{2}\right)}\right)}, & \thinspace & \left(a_{j},\beta_{l}\right)=0, & \thinspace & \left(d_{l},a_{j}\right)=w_{j}^{l}{\rm e}^{\eta_{j}},\\
 & \left(b_{j},\beta_{l}\right)=\frac{1+\mu_{j}}{2}\alpha_{j}^{\left(l\right)}, & \thinspace & \left(b_{j},\gamma_{l}\right)=0, & \thinspace & \left(d_{l},b_{j}\right)=0,\\
 & \left(a_{j},\gamma_{l}\right)=\frac{2ip_{l}r_{l}}{w_{j}+ip_{l}}, & \thinspace & \left(d_{l},\beta_{k}\right)=0, & \thinspace & \left(d_{l},\gamma_{l}\right)=\delta_{0l}\,r_{l},
\end{alignat*}
 where
\begin{alignat*}{2}
\eta_{j} =k_{j}x+\frac{\mu+\lambda}{k_{j}}t+\eta_{j,0},\qquad\varphi_{j} & =p_{j}x-\frac{\mu+\lambda}{p_{j}}t+\varphi_{j,0},
\end{alignat*}
and
\begin{gather*}
r^{2}=\sum\limits _{j=M_{1}+1}^{M}r_{j}^{2},\quad\mu_{j}=\begin{cases}
1 & 1\leq j\leq N\\
-1 & 1\leq j\leq N
\end{cases},\quad w_{j}=\begin{cases}
k_{j} & 1\leq j\leq N\\
k_{j}^{*} & N+1\leq j\leq2N
\end{cases}.
\end{gather*}
\end{thm}
The proof is presented in the Appendix.

\section{Conclusions}

In this paper, based on Hirota method, we have investigated the two- and multi-component AB system that describe the propagation of geophysical fluids. For the two-component system, we found the bright-dark one- and two-soliton solutions explicitly.
Asymptotic analysis  has been conducted on two-bright-dark-soliton solution to investigate the interactions between the two
solitons. We conclude that interactions between two bright or two dark solitons are both elastic. We should remark here that the interactions between two solitons for the multi-component integrable systems are non-elastic in some cases. 
We also illustrated  oblique interactions (Fig.\ref{fig:2sfc} and \ref{fig:2sdc}), bound states of solitons (Fig.\ref{fig:2sfp} and \ref{fig:2sdp}) analytically and graphically. The periods for bound states of solitons are given explicitly.  Furthermore, we give the $N$-bright-dark-soliton solutions to two- and $M$-component AB system in the form of Pfaffian.

\section*{\bf Acknowledgements}
G.F. Yu is supported by National Natural Science Foundation of
China (NSFC) (grant no.11371251) and Z.N. Zhu is supported by NSFC (grant nos.11428102 and 11671255), and by the Ministry of Economy and Competitiveness of Spain under contract MTM2016-80276-P (AEI/FEDER, EU). \vskip .5cm

\section{Appendix}
Proof of Theorem \ref{thm1}.\vskip  3pt
\begin{proof}
Because $\left(\bullet\right)=\left(a_{1},\cdots,a_{2N},b_{2N},\cdots,b_{1}\right)$
contains the index $a_{j}$, $\left(a_{j},d_{0},\bullet\right)$ is trivially
zero. We have
\begin{align}
\left(d_{k},a_{j}\right)\left(\bullet\right) & = \left(d_{k},a_{j}\right)\left(\bullet\right)+\left(a_{j},d_{k},\bullet\right)\nonumber \\
& = \sum_{l=1}^{2N}\left(-1\right)^{l}\left(a_{j},a_{l}\right)\left(d_{k},\hat{a}_{l}\right)+\left(-1\right)^{j+1}\left(a_{j},b_{j}\right)\left(d_{k},b_{j}\right).
\end{align}
Then
\begin{align}
 & g_{2}g_{2}^{*}-r^{2}f^{2}\nonumber \\
= & r^{2}\left(d_{0},\gamma,\bullet\right)\left(d_{0},\gamma^{*},\bullet\right)-r^{2}\left(\bullet\right)^{2}\nonumber \\
= & r^{2}\sum_{j=1}^{2N}\left(-1\right)^{j+1}\left[\left(\gamma,a_{j}\right)+\left(\gamma^{*},a_{j}\right)\right]\left(d_{0},\hat{a}_{j}\right)\left(\bullet\right)+r^{2}\sum_{j,l=1}^{2N}\left(-1\right)^{j+l}\left(\gamma,a_{j}\right)\left(\gamma^{*},a_{l}\right)\left(d_{0},\hat{a}_{j}\right)\left(d_{0},\hat{a}_{l}\right)\nonumber \\
= & r^{2}\sum_{j=1}^{2N}\left(-1\right)^{j}\frac{4p^{2}}{p^{2}+w_{j}^{2}}\left(d_{0},a_{j}\right)\left(d_{0},\hat{a}_{j}\right)\left(\bullet\right)+r^{2}\sum_{j,l=1}^{2N}\left(-1\right)^{j+l}\frac{4p^{2}}{\left(w_{j}+ip\right)\left(w_{l}-ip\right)}e^{\eta_{j}+\eta_{l}}\left(d_{0},\hat{a}_{j}\right)\left(d_{0},\hat{a}_{l}\right)\nonumber \\
= & r^{2}\sum_{j=1}^{2N}\left(-1\right)^{j}\frac{4p^{2}}{p^{2}+w_{j}^{2}}\left(-1\right)^{j+1}\left(d_{0},\hat{b}_{j}\right)
\left(d_{0},\hat{a}_{j}\right)+r^{2}\sum_{j=1}^{2N}\left[\left(-1\right)^{j}\frac{4p^{2}}{p^{2}+w_{j}^{2}}
\left(d_{0},\hat{a}_{j}\right)\sum_{l=1}^{2N}\left(-1\right)^{l}\left(a_{j},a_{l}\right)\left(d_{0},\hat{a}_{l}\right)\right]\nonumber \\& +r^{2}\sum_{j,l=1}^{2N}\left(-1\right)^{j+l}\frac{4p^{2}}{\left(w_{j}+ip\right)\left(w_{l}-ip\right)}e^{\eta_{j}+\eta_{l}}\left(d_{0},\hat{a}_{j}\right)\left(d_{0},\hat{a}_{l}\right)\nonumber \\
= & r^{2}\sum_{j=1}^{2N}-\frac{4p^{2}}{p^{2}+w_{j}^{2}}\left(d_{0},\hat{a}_{j}\right)\left(d_{0},\hat{b}_{j}\right)\nonumber \\
 & +2p^{2}r^{2}\sum_{j,l=1}^{2N}\left(-1\right)^{j+l}\left[\frac{\left(a_{j},a_{l}\right)}{p^{2}+w_{j}^{2}}+\frac{\left(a_{l},a_{j}\right)}{p^{2}+w_{l}^{2}}+\frac{e^{\eta_{j}+\eta_{l}}}{\left(w_{j}+ip\right)\left(w_{l}-ip\right)}+\frac{e^{\eta_{j}+\eta_{l}}}{\left(w_{l}+ip\right)\left(w_{j}-ip\right)}\right]\left(d_{0},\hat{a}_{j}\right)\left(d_{0},\hat{a}_{l}\right)\nonumber \notag
\end{align}
\begin{align}
= & -r^{2}\sum_{j=1}^{2N}\frac{4p^{2}}{p^{2}+w_{j}^{2}}\left(d_{0},\hat{a}_{j}\right)\left(d_{0},\hat{b}_{j}\right) +2p^{2}r^{2}\sum_{j,l=1}^{2N}\left(-1\right)^{j+l}\left(\frac{1}{p^{2}+w_{j}^{2}}+\frac{1}{p^{2}+w_{l}^{2}}\right)\left(d_{0},a_{j}\right)\left(d_{0},a_{l}\right)\left(d_{0},\hat{a}_{j}\right)\left(d_{0},\hat{a}_{l}\right)\nonumber \\
= & -4r^{2}p^{2}\sum_{j=1}^{2N}\frac{1}{p^{2}+w_{j}^{2}}\left(d_{0},\hat{a}_{j}\right)\left(d_{0},\hat{b}_{j}\right),
\end{align}
and
\begin{align}
g_{1}g_{1}^{*}= & \left(d_{0},\beta,\bullet\right)\left(d_{0},\beta^{*},\bullet\right)\nonumber \\
= & \sum_{j,l=1}^{2N}\left(-1\right)^{j+l}\frac{1+\mu_{j}}{2}\frac{1-\mu_{l}}{2}\left(d_{0},\hat{b}_{j}\right)\left(d_{0},\hat{b}_{l}\right)\nonumber \\
= & \sum_{j,l=1}^{2N}\left(-1\right)^{j+l}\frac{1-\mu_{j}\mu_{l}}{4}\left(d_{0},\hat{b}_{j}\right)\left(d_{0},\hat{b}_{l}\right)\nonumber \\
= & \frac{1}{2}\sum_{j,l=1}^{2N}\left(-1\right)^{j+l}\left(w_{j}^{2}-w_{l}^{2}\right)\left(\frac{2}{\sigma}+\frac{4p^{2}r^{2}}{\left(p^{2}+w_{j}^{2}\right)\left(p^{2}+w_{l}^{2}\right)}\right)\left(b_{j},b_{l}\right)\left(d_{0},\hat{b}_{j}\right)\left(d_{0},\hat{b}_{l}\right)\nonumber \\
= & \frac{2}{\sigma}\sum_{j=1}^{2N}\left[\left(-1\right)^{j+1}w_{j}^{2}\left(d_{0},\hat{b}_{j}\right)\sum_{l=1}^{2N}\left(-1\right)^{l+1}\left(b_{j},b_{l}\right)\left(d_{0},\hat{b}_{l}\right)\right]\notag\\
&\quad+2p^{2}r^{2}\sum_{j,l=1}^{2N}\left(-1\right)^{j+l}\left(\frac{1}{p^{2}+w_{l}^{2}}-\frac{1}{p^{2}+w_{j}^{2}}\right)\left(b_{j},b_{l}\right)\left(d_{0},\hat{b}_{j}\right)\left(d_{0},\hat{b}_{l}\right)\nonumber \\
= & \frac{2}{\sigma}\sum_{j=1}^{2N}\left[\left(-1\right)^{j+1}w_{j}^{2}\left(d_{0},\hat{b}_{j}\right)\left(\left(b_{j},d_{0},\bullet\right)-\left(-1\right)^{j}\left(b_{j},a_{j}\right)\left(d_{0},\hat{a}_{j}\right)\right)\right]\nonumber \\
 & +4p^{2}r^{2}\left[\sum_{j=1}^{2N}\left(-1\right)^{j}\frac{1}{p^{2}+w_{j}^{2}}\left(d_{0},\hat{b}_{j}\right)\sum_{l=1}^{2N}\left(-1\right)^{l+1}\left(b_{j},b_{l}\right)\left(d_{0},\hat{b}_{l}\right)\right]\nonumber \\
= & -\frac{2}{\sigma}\sum_{j=1}^{2N}w_{j}^{2}\left(d_{0},\hat{a}_{j}\right)\left(d_{0},\hat{b}_{j}\right)
+4p^{2}r^{2}\sum_{j=1}^{2N}\frac{1}{p^{2}+w_{j}^{2}}\left(d_{0},\hat{a}_{j}\right)\left(d_{0},\hat{b}_{j}\right),
\end{align}

and
\begin{align}
-\left(f_{x}\right)^{2}= & \left(d_{2},d_{0},\bullet\right)\left(d_{0},d_{0},\bullet\right)-\left(d_{1},d_{0},\bullet\right)^{2}\nonumber \\
= & \sum_{j,l=1}^{2N}\left(-1\right)^{j+l}\left[\left(d_{2},a_{j}\right)\left(d_{0},a_{l}\right)-\left(d_{1},a_{j}\right)\left(d_{1},a_{l}\right)\right]\left(d_{0},\hat{a}_{j}\right)\left(d_{0},\hat{a}_{l}\right)\nonumber \\
= & \sum_{j,l=1}^{2N}\left(-1\right)^{j+l}\left(\frac{w_{j}^{2}+w_{l}^{2}}{2}-w_{j}w_{l}\right)e^{\eta_{j}+\eta_{l}}\left(d_{0},\hat{a}_{j}\right)\left(d_{0},\hat{a}_{l}\right)\nonumber \\
= & \sum_{j,l=1}^{2N}\left(-1\right)^{j+l}\frac{w_{l}^{2}-w_{j}^{2}}{2}\left(a_{j},a_{l}\right)\left(d_{0},\hat{a}_{j}\right)\left(d_{0},\hat{a}_{l}\right)\nonumber \\
= & \sum_{j=1}^{2N}\left[\left(-1\right)^{j}\left(-w_{j}^{2}\right)\left(d_{0},\hat{a}_{j}\right)\sum_{l=1}^{2N}\left(-1\right)^{l}\left(a_{j},a_{l}\right)\left(d_{0},\hat{a}_{l}\right)\right]\nonumber \notag
\end{align}
\begin{align}
= & -\sum_{j=1}^{2N}\left(-1\right)^{j}\left[w_{j}^{2}\left(d_{0},a_{j}\right)\right]\left(d_{0},\hat{a}_{j}\right)\left(\bullet\right)-\sum_{j=1}^{2N}w_{j}^{2}\left(a_{j},b_{j}\right)\left(d_{0},\hat{a}_{j}\right)\left(d_{0},\hat{b}_{j}\right)\nonumber \\
= & -\left(d_{2},d_{0},\bullet\right)\left(\bullet\right)-\sum_{j=1}^{2N}w_{j}^{2}\left(d_{0},\hat{a}_{j}\right)\left(d_{0},\hat{b}_{j}\right)\nonumber \\
= & -f_{xx}f+\frac{\sigma}{2}\left(g_{1}g_{1}^{*}+g_{2}g_{2}^{*}-r^{2}f^{2}\right).
\end{align}
Substitution above expressions into eq.(\ref{eq:b2c3}) leads to all terms cancelled.
Thus eq. (\ref{eq:b2c3}) holds and the proof is completed.
\end{proof}

Proof of Theorem \ref{thm2}.

\begin{proof}
Upon properties of Pfaffian, we can derive that
\begin{align*}
 & f_{x}=\left(d_{1},d_{0},\bullet\right),\quad f_{t}=\left(\lambda+\mu\right)\left(d_{0},d_{-1},\bullet\right),\\
 & f_{xx}=\left(d_{2},d_{0},\bullet\right),\quad f_{xt}=\left(\lambda+\mu\right)\left(d_{1},d_{-1},\bullet\right),\\
 & g_{j,x}=\left(d_{1},\beta_{j},\bullet\right),\quad g_{j,t}=\left(\lambda+\mu\right)\left(d_{-1},\beta_{j},\bullet\right),\\
 & g_{j,xt}=\left(\lambda+\mu\right)\left[\left(d_{0},\beta_{j},\bullet\right)+\left(d_{1},d_{0},d_{-1},\beta_{j},\bullet\right)\right],\\
 & \partial_{x}\left(d_{0},\gamma_{j},\bullet\right)=ip_{j}r_{j}\left(\bullet\right)-ip_{j}\left(d_{0},\gamma_{j},\bullet\right)+r_{j}\left(d_{1},d_{0},\bullet\right)-\left(d_{1},\gamma_{j},\bullet\right),\\
 & \partial_{t}\left(d_{0},\gamma_{j},\bullet\right)=(\lambda+\mu)\Big[\frac{r_{j}}{ip_{j}}\left(\bullet\right)
 -\frac{1}{ip_{j}}\left(d_{0},\gamma_{j},\bullet\right)
 -r_{j}\left(d_{0},d_{-1},\bullet\right)+\left(d_{-1},\gamma_{j},\bullet\right)\Big],\\
 & \partial_{x}\partial_t\left(d_{0},\gamma_{j},\bullet\right)=\left(\lambda+\mu\right)\Big[2\left(d_{0},\gamma_{j},\bullet\right)-2r_{j}\left(\bullet\right)+2ip_{j}r_{j}\left(d_{0},d_{-1},\bullet\right)\\
 & \qquad-ip_{j}\left(d_{-1},\gamma_{j},\bullet\right)-\frac{1}{ip_{j}}\left(d_{1},\gamma_{j},\bullet\right)
 +\left(d_{1},d_{0},d_{-1},\gamma_{j},\bullet\right)\Big].
\end{align*}
Substituting above expression into  eq.(\ref{eq:bim1}) directly, we arrive at the Pfaffian identity. Thus the Pfaffian expression solve  eq.(\ref{eq:bim1}). To prove eq.(\ref{eq:bim2}), we introduce additional Pfaffian entries
\begin{alignat*}{3}
 & \left(b_{j},\beta_{l}^{*}\right)=\frac{1-\mu_{j}}{2}\alpha_{l}^{\left(n\right)}, & \thinspace & \left(a_{j},\beta_{l}^{*}\right)=0, & \thinspace & \left(d_{0},\beta^{*}\right)=0,\\
 & \left(a_{j},\gamma_{l}^{*}\right)=-\frac{2ipr_{j}}{w_{j}-ip}, & \thinspace & \left(b_{j},\gamma_{l}^{*}\right)=0, & \thinspace & \left(d_{0},\gamma^{*}\right)=1,
\end{alignat*}
such that we can present the Pfaffian form for $g_{j}^{*}$ as
\begin{align}
g_{j}^{*} & = \left(d_{0},\beta_{j},a_{1},\cdots,a_{2N},b_{2N},\cdots,b_{1}\right)=\left(d_{0},\beta_{j},\bullet\right),\thinspace j=1,\cdots,M_{1},\\
g_{j}^{*} & =  e^{-i\varphi_{j}}\left(d_{0},\gamma_{j},a_{1},\cdots,a_{2N},b_{2N},\cdots,b_{1}\right)=e^{-i\varphi_{j}}\left(d_{0},\gamma_{j},\bullet\right),\thinspace j=M_{1}+1,\cdots,M.
\end{align}
Then we have
\begin{align*}
 & \sum_{j=M_{1}+1}^{M}\left(d_{0},\gamma_{j},\bullet\right)\left(d_{0},\gamma_{j}^{*},\bullet\right)-r^{2}\left(\bullet\right)^{2}\\
= & \sum_{n=M_{1}+1}^{M}r_{n}\left[\sum_{j=1}^{2N}\left(-1\right)^{j+1}\left[\left(\gamma_{n},a_{j}\right)+\left(\gamma_{n}^{*},a_{j}\right)\right]\left(d_{0},\hat{a}_{j}\right)\left(\bullet\right)\right]+\sum_{n=M_{1}+1}^{M}\sum_{j,l=1}^{2N}\left(-1\right)^{j+l}\left(\gamma_{n},a_{j}\right)\left(\gamma_{n}^{*},a_{l}\right)\left(d_{0},\hat{a}_{j}\right)\left(d_{0},\hat{a}_{l}\right)\\
= & \sum_{n=M_{1}+1}^{M}r_{n}^{2}\sum_{j=1}^{2N}\left(-1\right)^{j}\frac{4p_{n}^{2}}{p_{n}^{2}+w_{j}^{2}}\left(d_{0},a_{j}\right)\left(d_{0},\hat{a}_{j}\right)\left(\bullet\right)+\sum_{n=M_{1}+1}^{M}r_{n}^{2}\sum_{j,l=1}^{2N}\left(-1\right)^{j+l}\frac{4p_{n}^{2}}{\left(w_{j}+ip_{n}\right)\left(w_{l}-ip_{n}\right)}e^{\eta_{j}+\eta_{l}}\left(d_{0},\hat{a}_{j}\right)\left(d_{0},\hat{a}_{l}\right)\\
\end{align*}
\begin{align*}
= & \sum_{n=M_{1}+1}^{M}r_{n}^{2}\sum_{j=1}^{2N}\left(-1\right)^{j}\frac{4p_{n}^{2}}{p_{n}^{2}+w_{j}^{2}}\left(-1\right)^{j+1}\left(d_{0},\hat{b}_{j}\right)\left(d_{0},\hat{a}_{j}\right)+\sum_{n=M_{1}+1}^{M}r_{n}^{2}\sum_{j=1}^{2N}\left[\left(-1\right)^{j}\frac{4p_{n}^{2}}{p_{n}^{2}+w_{j}^{2}}\left(d_{0},\hat{a}_{j}\right)\sum_{l=1}^{2N}\left(-1\right)^{l}\left(a_{j},a_{l}\right)\left(d_{0},\hat{a}_{l}\right)\right]\\
 & +\sum_{n=M_{1}+1}^{M}r_{n}^{2}\sum_{j,l=1}^{2N}\left(-1\right)^{j+l}\frac{4p_{n}^{2}}{\left(w_{j}+ip_{n}\right)\left(w_{l}-ip_{n}\right)}e^{\eta_{j}+\eta_{l}}\left(d_{0},\hat{a}_{j}\right)\left(d_{0},\hat{a}_{l}\right)\\
= & \sum_{n=M_{1}+1}^{M}r_{n}^{2}\sum_{j=1}^{2N}-\frac{4p_{n}^{2}}{p_{n}^{2}+w_{j}^{2}}\left(d_{0},\hat{a}_{j}\right)\left(d_{0},\hat{b}_{j}\right)\\
 & +\sum_{n=M_{1}+1}^{M}2r_{n}^{2}p_{n}^{2}\sum_{j,l=1}^{2N}\left(-1\right)^{j+l}\left[\frac{\left(a_{j},a_{l}\right)}{p_{n}^{2}+w_{j}^{2}}+\frac{\left(a_{l},a_{j}\right)}{p_{n}^{2}+w_{l}^{2}}+\frac{e^{\eta_{j}+\eta_{l}}}{\left(w_{j}+ip_{n}\right)\left(w_{l}-ip_{n}\right)}+\frac{e^{\eta_{j}+\eta_{l}}}{\left(w_{l}+ip_{n}\right)\left(w_{j}-ip_{n}\right)}\right]\left(d_{0},\hat{a}_{j}\right)\left(d_{0},\hat{a}_{l}\right)\\
= & -\sum_{n=M_{1}+1}^{M}r_{n}^{2}\sum_{j=1}^{2N}\frac{4p_{n}^{2}}{p_{n}^{2}+w_{j}^{2}}\left(d_{0},\hat{a}_{j}\right)\left(d_{0},\hat{b}_{j}\right)\\
 & +\sum_{n=M_{1}+1}^{M}2r_{n}^{2}p_{n}^{2}\sum_{j,l=1}^{2N}\left(-1\right)^{j+l}\left(\frac{1}{p_{n}^{2}+w_{j}^{2}}+\frac{1}{p_{n}^{2}+w_{l}^{2}}\right)\left(d_{0},a_{j}\right)\left(d_{0},a_{l}\right)\left(d_{0},\hat{a}_{j}\right)\left(d_{0},\hat{a}_{l}\right)\\
= & -\sum_{n=M_{1}+1}^{M}\sum_{j=1}^{2N}\frac{4r_{n}^{2}p_{n}^{2}}{p_{n}^{2}+w_{j}^{2}}\left(d_{0},\hat{a}_{j}\right)\left(d_{0},\hat{b}_{j}\right),
\end{align*}
and
\begin{align*}
 & \sigma\sum_{j=1}^{M_{1}}\left(d_{0},\beta_{j},\bullet\right)\left(d_{0},\beta_{j}^{*},\bullet\right)\\
= & \sigma\sum_{n=1}^{M_{1}}\sum_{j,l=1}^{2N}\left(-1\right)^{j+l}\frac{1+\mu_{j}}{2}\frac{1-\mu_{l}}{2}\alpha_{j}^{\left(n\right)}\alpha_{l}^{\left(n\right)}\left(d_{0},\hat{b}_{j}\right)\left(d_{0},\hat{b}_{l}\right)\\
= & \sigma\sum_{j,l=1}^{2N}\left(-1\right)^{j+l}\frac{1-\mu_{j}\mu_{l}}{4}\left[\sum_{n=1}^{M_{1}}\alpha_{j}^{\left(n\right)}\alpha_{l}^{\left(n\right)}\right]\left(d_{0},\hat{b}_{j}\right)\left(d_{0},\hat{b}_{l}\right)\\
= & \frac{\sigma}{2}\sum_{j,l=1}^{2N}\left(-1\right)^{j+l}\left(w_{j}^{2}-w_{l}^{2}\right)\left(\frac{2}{\sigma}+\sum_{n=M_{1}+1}^{M}\frac{4p_{n}^{2}r_{n}^{2}}{\left(p_{n}^{2}+w_{j}^{2}\right)\left(p_{n}^{2}+w_{l}^{2}\right)}\right)\left(b_{j},b_{l}\right)\left(d_{0},\hat{b}_{j}\right)\left(d_{0},\hat{b}_{l}\right)
\end{align*}
\begin{align*}
= & 2\sum_{j=1}^{2N}\left[\left(-1\right)^{j+1}w_{j}^{2}\left(d_{0},\hat{b}_{j}\right)\sum_{l=1}^{2N}\left(-1\right)^{l+1}\left(b_{j},b_{l}\right)\left(d_{0},\hat{b}_{l}\right)\right]\\
 & +\sum_{n=M_{1}+1}^{M}2\sigma p_{n}^{2}r_{n}^{2}\sum_{j,l=1}^{2N}\left(-1\right)^{j+l}\left(\frac{1}{p_{n}^{2}+w_{l}^{2}}-\frac{1}{p_{n}^{2}+w_{j}^{2}}\right)\left(b_{j},b_{l}\right)\left(d_{0},\hat{b}_{j}\right)\left(d_{0},\hat{b}_{l}\right)\\
= & 2\sum_{j=1}^{2N}\left[\left(-1\right)^{j+1}w_{j}^{2}\left(d_{0},\hat{b}_{j}\right)\left(\left(b_{j},d_{0},\bullet\right)-\left(-1\right)^{j}\left(b_{j},a_{j}\right)\left(d_{0},\hat{a}_{j}\right)\right)\right]\\
 & +\sum_{n=M_{1}+1}^{M}4\sigma p_{n}^{2}r_{n}^{2}\left[\sum_{j=1}^{2N}\left(-1\right)^{j}\frac{1}{p_{n}^{2}+w_{j}^{2}}\left(d_{0},\hat{b}_{j}\right)\sum_{l=1}^{2N}\left(-1\right)^{l+1}\left(b_{j},b_{l}\right)\left(d_{0},\hat{b}_{l}\right)\right]\\
= & -2\sum_{j=1}^{2N}w_{j}^{2}\left(d_{0},\hat{a}_{j}\right)\left(d_{0},\hat{b}_{j}\right)+\sum_{n=M_{1}+1}^{M}4\sigma p_{n}^{2}r_{n}^{2}\sum_{j=1}^{2N}\frac{1}{p_{n}^{2}+w_{j}^{2}}\left(d_{0},\hat{a}_{j}\right)\left(d_{0},\hat{b}_{j}\right),
\end{align*}

and
\begin{align*}
-\left(f_{x}\right)^{2}= & \left(d_{2},d_{0},\bullet\right)\left(d_{0},d_{0},\bullet\right)-\left(d_{1},d_{0},\bullet\right)^{2}\\
= & \sum_{j,l=1}^{2N}\left(-1\right)^{j+l}\left[\left(d_{2},a_{j}\right)\left(d_{0},a_{l}\right)-\left(d_{1},a_{j}\right)\left(d_{1},a_{l}\right)\right]\left(d_{0},\hat{a}_{j}\right)\left(d_{0},\hat{a}_{l}\right)\\
= & \sum_{j,l=1}^{2N}\left(-1\right)^{j+l}\left(\frac{w_{j}^{2}+w_{l}^{2}}{2}-w_{j}w_{l}\right)e^{\eta_{j}+\eta_{l}}\left(d_{0},\hat{a}_{j}\right)\left(d_{0},\hat{a}_{l}\right)\\
= & \sum_{j,l=1}^{2N}\left(-1\right)^{j+l}\frac{w_{l}^{2}-w_{j}^{2}}{2}\left(a_{j},a_{l}\right)\left(d_{0},\hat{a}_{j}\right)\left(d_{0},\hat{a}_{l}\right)\\
= & \sum_{j=1}^{2N}\left[\left(-1\right)^{j}\left(-w_{j}^{2}\right)\left(d_{0},\hat{a}_{j}\right)\sum_{l=1}^{2N}\left(-1\right)^{l}\left(a_{j},a_{l}\right)\left(d_{0},\hat{a}_{l}\right)\right]\\
= & -\sum_{j=1}^{2N}\left(-1\right)^{j}\left[w_{j}^{2}\left(d_{0},a_{j}\right)\right]\left(d_{0},\hat{a}_{j}\right)\left(\bullet\right)-\sum_{j=1}^{2N}w_{j}^{2}\left(a_{j},b_{j}\right)\left(d_{0},\hat{a}_{j}\right)\left(d_{0},\hat{b}_{j}\right)\\
= & -\left(d_{2},d_{0},\bullet\right)\left(\bullet\right)-\sum_{j=1}^{2N}w_{j}^{2}\left(d_{0},\hat{a}_{j}\right)\left(d_{0},\hat{b}_{j}\right).
\end{align*}
One can check that when substitute above identities into eq.(\ref{eq:bim2}), all terms are cancelled.
Thus we have proved \eqref{mn-1}-\eqref{mn-3} solve eq.(\ref{eq:bim2}).
\end{proof}

\end{document}